\documentclass[a4paper,11pt,final]{article}

\usepackage[colorlinks=true]{hyperref}
\usepackage[english]{babel}
\usepackage[utf8]{inputenc}
\usepackage[T1]{fontenc}
\usepackage{lmodern}
\usepackage[pdftex]{graphicx}
\usepackage[top=2cm, bottom=2cm, left=2cm, right=2cm]{geometry}
\usepackage{setspace}
\usepackage[french]{varioref}
\usepackage{lastpage}
\usepackage{fancyhdr}
\usepackage[table]{xcolor}
\usepackage{tikz}
\usepackage{titling}
\usepackage{graphicx}
\usepackage{amsmath}
\usepackage{amsthm}
\usepackage{amssymb}

\usepackage{multirow}
\usepackage{rotating}
\usepackage{subcaption} 
\usepackage[gen]{eurosym}
\usepackage{dirtree}

\usepackage{rotating}
\usepackage{colortbl}

\usepackage{tabularx}

\newcolumntype{C}[1]{>{\centering\let\newline\\\arraybackslash\hspace{0pt}}m{#1}}
\newcolumntype{L}[1]{>{\raggedright\let\newline\\\arraybackslash\hspace{0pt}}m{#1}}
\newcolumntype{R}[1]{>{\raggedleft\let\newline\\\arraybackslash\hspace{0pt}}m{#1}}

\usepackage{doi}
\usepackage[autostyle]{csquotes}
\usepackage[style=numeric, citestyle=numeric-comp, 
	maxcitenames=2, maxbibnames=99, 
    backend=biber
]{biblatex}
\usepackage
	{todonotes}

\title{Multiple Partitioning of Multiplex Signed Networks: \\ Application to European Parliament Votes}
\author{Nejat Arinik, Rosa Figueiredo \& Vincent Labatut}

\hypersetup{
    pdftitle={\thetitle},
    pdfauthor={\theauthor},
    bookmarksnumbered=true,bookmarksopen=true,
	unicode=true,colorlinks=true,linktoc=all,
	linkcolor=blue,citecolor=blue,filecolor=blue,urlcolor=blue,
	pdfstartview=FitH
}

\usepackage{enumitem}
\setlist{nolistsep}

\begin{document}
\maketitle
\sloppy

\abstract{For more than a decade, graphs have been used to model the voting behavior taking place in parliaments. However, the methods described in the literature suffer from several limitations. The two main ones are that 1) they rely on some temporal integration of the raw data, which causes some information loss; and/or 2) they identify groups of antagonistic voters, but not the context associated to their occurrence. In this article, we propose a novel method taking advantage of multiplex signed graphs to solve both these issues. It consists in first partitioning separately each layer, before grouping these partitions by similarity. We show the interest of our approach by applying it to a European Parliament dataset.}


\textcolor{red}{\textbf{Cite as:} N. Arinik, R. Figueiredo \& V. Labatut. \href{https://www.sciencedirect.com/science/article/pii/S0378873318300303}{Multiple Partitioning of Multiplex Signed Networks: Application to European Parliament Votes}. Social Networks, 2020, 60:83-102. DOI: \href{www.doi.org/10.1016/j.socnet.2019.02.001}{10.1016/j.socnet.2019.02.001}}

\section{Introduction}
\label{sec:Intro}





In a \textit{signed graph}, each link is associated to a sign, which can be either positive ($+$) or negative ($-$). This type of graph was originally introduced in Psychology, as a means to describe relationships between people belonging to distinct social groups~\cite{Heider1946}. More generally, they can be used to model any system containing two types of antithetical relationships (like/dislike, for/against, similar/different...). A signed graph is considered \textit{structurally balanced} if it can be partitioned into two~\cite{Cartwright1956} or more~\cite{Davis1967} clusters, such that positive links are located inside the clusters, and negatives ones are in-between them. For instance, in the case of a social network whose links represent like/dislike relationships, this amounts to having mutually hostile social groups with internal friendship.

However, it is very rare for a real-world network to have a perfectly balanced structure: the question is then to quantify how imbalanced it is. Various measures have been defined for this purpose, the simplest consisting in counting the numbers of misplaced links, i.e. negative ones located inside the groups, and positive ones located between them \cite{Cartwright1956}. Such measures are expressed relatively to a graph partition, so processing the graph balance amounts to identifying the partition corresponding to the lowest imbalance measure. In other words, calculating the graph balance can be formulated as an optimization problem. This type of optimization problem can be compared to that of \textit{community detection}, which consists in partitioning \textit{unsigned} networks in order to detect groups of nodes more densely connected relatively to the rest of the network \cite{Fortunato2010}. The main difference is of course the presence of signs attached to links, which represent additional information one has to take into account. Doing so is a non-trivial task, which cannot be conducted by simply performing minor adaptations of community detection methods \cite{Chiang2014}. 

Signed graph partitioning is also an important problem in terms of applications: it is used in a number of situations to get a better understanding of the studied real-world system, be it social \cite{Doreian1996}, biological \cite{DasGupta2007}, diplomatic \cite{Estrada2014b}, business-related \cite{Jensen2006}, sports-related \cite{Kaplan2008}, judicial \cite{Mrvar2009}, bibliographic \cite{Huang2010a}, political \cite{Cucuringu2013}, conversational \cite{Hassan2012a}, financial \cite{MacMahon2013}, etc. It is also used to solve some problems of interest related to the considered systems, e.g. portfolio optimization \cite{MacMahon2013}, opinion group detection in online discussions \cite{Hassan2012a}, decomposition of biological systems \cite{DasGupta2007}, document classification \cite{Bansal2002}, surface detection in 3D images \cite{Kolluri2004}, or detection of embedded matrix structures \cite{Figueiredo2011}.

In this work, our goal is to use this paradigm to study the voting activity of the Members of the European Parliament (MEPs). We want not only to detect groups of MEPs which would be cohesive in terms of votes, but also to identify the different typical \textit{voting behavior patterns} of the European Parliament (EP), i.e. the characteristic ways in which the MEP set is partitioned by these votes. The publicly available raw data is basically a table showing how each MEP voted at each roll-call. As we will see later in the bibliographical section, the standard approach to study this type of system is to extract a vote similarity network, in which nodes represent MEPs, and weighted, possibly signed, links represent the similarity between two MEPs, averaged over the series of roll-calls (e.g. \cite{Waugh2011, Traag2013, Mendonca2015}). However, this averaging leads to some information loss due to the temporal integration performed on the raw data. Moreover, as shown in \cite{Arinik2017}, this approach does not allow identifying the typical voting patterns of the parliament.

Instead, we propose to adopt an approach based on a multiplex signed vote similarity network, in which each layer models a single roll-call as a signed unweighted graph. The literature contains $3$ main approaches to partition multiplex graphs in general \cite{Kivelae2013}: 1) merge the layers and apply a traditional partitioning method to the resulting aggregated graph; 2) apply a traditional method separately to each layer and merge the resulting partitions; and 3) use a method specifically designed for multilayer graphs, which partitions the set of all nodes (over all layers). All $3$ approaches are based on the assumption that one is looking for a \textit{single} partition. In both first approaches, each cluster of the partition contains all instances of the same node over all layers: it can be considered as a consensual partition, fitting all layers at once. In the latter approach, a community does not necessarily span all layers (two instances of the same node can belong to different communities). This also holds for multiplex \textit{signed} networks, in which certain algorithms look for a consensual partition \cite{Cucuringu2013, Chang2013b, Zhang2013r,  Li2016s, Zhan2016a}, while others search a trans-layer one \cite{Mucha2010a}.

This single-partition assumption is not compatible with all our objectives. Indeed, we look for the \textit{different} typical voting patterns of the EP, so we want our method to be able to identify \textit{several} partitions. To this aim, we propose a new partitioning process for multiplex signed graphs, which is four-stepped. First, we separately partition each layer, through a standard signed graph partitioning method. Second, we compute the similarity between all pairs of the resulting partitions, and third we perform a standard cluster analysis in order to identify clusters of similar partitions. Fourth, we process a characteristic partition for each cluster, which can be used for interpretation purposes. We apply our method to a dataset representing the voting activity during the 7\textsuperscript{th} term of the EP. As desired, it allows identifying groups of cohesive voters, but also their different characteristic voting configurations, as well as the legislative propositions to which they apply. In particular, we focus on certain case studies previously analyzed and discussed by Arinik \textit{et al}. \cite{Arinik2017} using a standard approach, and show that our method not only confirms certain of their assumptions, but also uncovers previously overlooked properties.

The rest of the article is organized as follows. First, in Section~\ref{sec:RelatedWork}, we review the previous works dealing with the extraction and analysis of vote networks representing parliaments and related institutions. In Section~\ref{sec:Dataset}, we then describe how the dataset that we use was constituted, as well as its most relevant properties. We turn to the methods in Section~\ref{sec:Methods}, and describe the approach we propose for the analysis of multiplex signed networks. In Section~\ref{sec:Results}, we present our results on a few specific cases selected from the dataset, and discuss them. Finally, in Section~\ref{sec:Conclusion} we summarize our findings, comment the limitations of our work and describe how they can be overcome, and how our work can be extended.

\section{Related Work}
\label{sec:RelatedWork}
Although the voting behavior in parliaments and related institutions has been studied for decades \cite{Attina1990}, the use of graphs as a modeling tool is quite recent, probably concomitant to the emergence of appropriate graph analysis methods. Table~\ref{tab:Biblio} lists the works related to this topic, with their main characteristics. It shows that only a small number of institutions have been studied up to now: United Nation General Assembly (UNGA), US Congress (or sometimes only the House of Representatives or Senate), US Supreme Court, European Parliament (EP), Brazilian and Italian Houses of Deputies.

\begin{table}[htb]
	\caption{Articles related to the analysis of voting data through graph-based models, which can be weighted (W), signed (S), dynamic (D), and bipartite (B).}
    \label{tab:Biblio}
	\centering
	\begin{tabular}{l l r c c c c}
    	\hline
        References & Institution & Period & W & S & D & B \\
        \hline
        Mrvar \& Doreian, 2009 \cite{Mrvar2009} & US Supreme Court & 2006--2007 &  & $\checkmark$ &  & $\checkmark$ \\
        Waugh \textit{et al}., 2009 \cite{Waugh2011} & US Congress & 1789--2007 & $\checkmark$ &  &  &  \\
        Balasubramanyan \textit{et al}., 2010 \cite{Balasubramanyan2010} & US Senate & N/A & $\checkmark$ &  &  &  \\
        Mucha \textit{et al}., 2010 \cite{Mucha2010,Mucha2010a} & US Senate & 1789--2007 & $\checkmark$ &  & $\checkmark$ &  \\
        Guimerà \& Sales-Pardo, 2011 \cite{Guimera2011} & US Supreme Court & 1953-–2004 &  & $\checkmark$ &  & $\checkmark$ \\
        Reda \textit{et al}., 2011 \cite{Reda2011} & US Representatives & 2010 &  &  & $\checkmark$ &  \\
        Macon \textit{et al}., 2012 \cite{Macon2012} (1) & UNGA & 1946--2008 & $\checkmark$ &  &  &  \\
        Macon \textit{et al}., 2012 \cite{Macon2012} (2) & UNGA & 1946--2008 & $\checkmark$ & $\checkmark$ &  &  \\
        Macon \textit{et al}., 2012 \cite{Macon2012} (3) & UNGA & 1946--2008 &  & $\checkmark$ &  & $\checkmark$ \\
        Wang \textit{et al}., 2012 \cite{Wang2012aa} & US Representatives & 2008-2010 & \multicolumn{4}{c}{Heterogeneous} \\
        Doreian \textit{et al}., 2013 \cite{Doreian2013} & UNGA & 1981--2001 &  & $\checkmark$ &  & $\checkmark$ \\
        Traag \textit{et al}., 2013 \cite{Traag2013} & European Parliament & 1979--2009 & $\checkmark$ &  &  &  \\
        Figueiredo \& Frota, 2014 \cite{Figueiredo2014} & UNGA & 1946--2008 &  & $\checkmark$ &  &  \\
        Dal Maso \textit{et al}., 2014 \cite{Maso2014} & Italian Deputies & 2013 & $\checkmark$ &  &  &  \\
        West \textit{et al}., 2014 \cite{West2014} & US Representatives & 2005 &  & $\checkmark$ &  &  \\
        Mendon\c{c}a \& Arinik \textit{et al}., 2015 \cite{Mendonca2015,Arinik2017} & European Parliament & 2009-2014 & $\checkmark$ & $\checkmark$ &  &  \\
        Santamaria \& Gomez, 2015 \cite{Santamaria2015} & European Parliament & 2014--2015 & $\checkmark$ & $\checkmark$ &  &  \\
        Crane \& Dempsey, 2016 \cite{Crane2016} & US Congress & 2001--2003 & $\checkmark$ &  &  &  \\
        Levorato \& Frota, 2017 \cite{Levorato2017} & Brazilian Deputies & 2011--2016 & $\checkmark$ & $\checkmark$ &  &  \\
        Levorato \textit{et al}., 2017 \cite{Levorato2017a} & UNGA & 1946--2008 & $\checkmark$ & $\checkmark$ &  &  \\
        \hline
	\end{tabular}
\end{table}

The nature of the analysis performed on the graph mainly depends on how the network is extracted from the raw vote data. The considered institutions function in very similar ways, so the methods used to perform this extraction do not change much from one author to the other. One can mainly distinguish two approaches: vote similarity networks, and agent-bill networks.


\subsection{Vote Similarity Networks}
\label{sec:RelWorkVoteSimNet}
Most of the works from the literature use what we call \textit{vote similarity networks} to model the voting activity. The methods to extract these networks are based on the same general principle, but can differ on a few points: the way they handle absence, abstention, time, and the possible filtering of the raw data and/or the obtained networks. For this reason, instead of considering each work separately, we prefer to group them according to their methodological choices in the following description. 

In a vote similarity network, nodes represent voting agents (such as legislators), and links model the similarity between them, relatively to the way they vote. These networks are most of the time weighted: a \textit{positive} numerical value associated to a link represents the level of similarity between the nodes it connects. Such a weight is processed thanks to a \textit{similarity function}, whose role is to estimate the level of agreement between two agents of interest, based on the votes they cast during a selection of roll-calls. This is an important methodological choice, because the resulting network is obtained by \textit{temporal integration} of the raw data (collection of votes). 

The basic approach consists in computing the number of exactly similar votes (a.k.a. \textit{full agreement}): \textsc{For}-\textsc{For} and \textsc{Against}-\textsc{Against} \cite{Macon2012}. But most authors normalize this value using the number of considered roll-calls, in order to get a proportion \cite{Waugh2011, Traag2013, Maso2014, Crane2016}. The resulting network is fully connected, i.e. there is a weighted link between each pair of nodes. Some authors prefer to extract \textit{signed} vote similarity networks, which means that a weight can be positive, but also negative \cite{Macon2012, Figueiredo2014, Mendonca2015, Levorato2017, Arinik2017, Levorato2017a}. In their similarity function, they count agreement positively and disagreement negatively. 

Depending on the modeled institution, it is generally possible for the agents to be signaled as \textsc{Absent}, which means they did not vote at all. This affects the similarity function in two ways in the literature. On the one hand, certain authors simply ignore the cases where at least one of the two considered agents was absent, when processing the proportion of agreement \cite{Waugh2011, Traag2013, Maso2014}: the available information is just insufficient to determine if there was agreement. On the other hand, some authors prefer to take them into account \cite{Mendonca2015, Levorato2017, Arinik2017}, because ignoring them gives much importance to the few votes cast by chronically absent agents.

In most institutions, the agents can also explicitly cast an \textsc{Abstain} vote, allowing them to express their will to not take any position regarding the issue at hand. This is the point on which authors differ the most when defining their similarity functions. Traag \textit{et al}. treat abstention as absence \cite{Traag2013}, considering the available information is insufficient to infer agreement or disagreement. With unsigned approaches, the rest of the options are to consider abstention either as disagreement ($0$), agreement ($+1$), or half-agreement ($+0.5$). By comparison, disagreement is modeled with a negative value ($-1$) in signed similarity functions, so they allow properly treating abstention as a distinct case ($0$). The literature contains various combinations of these options. Certain authors only consider \textsc{Abstain}-\textsc{Abstain} as agreement, and treat all other abstention situations like absences (by ignoring them) \cite{Waugh2011} or like disagreements \cite{Maso2014, Levorato2017}. Levorato \textit{et al}. consider \textsc{Abstain}-\textsc{Abstain} as full agreement ($+1$), whereas \textsc{Abstain}-\textsc{For} and \textsc{Abstain}-\textsc{Against} are half agreement ($+0.5$) \cite{Levorato2017a}. Reda \textit{et al}. and Macon \textit{et al}. consider all combinations of absences and abstentions (\textsc{Absent}-\textsc{Absent}, \textsc{Abstain}-\textsc{Abstain}, and \textsc{Absent}-\textsc{Abstain}) as agreement \cite{Reda2011}. In their signed approach, Mendon\c{c}a \textit{et al}. compare two variants of such approaches  \cite{Mendonca2015}: half agreement when at least one agent abstains, vs. no agreement ($0$) when there is exactly one abstention. They do not find any significant differences. Levorato \& Frota also compare these variants, and observe more variable effects. Moreover, they must deal with \textsc{Obstruction}, a type of vote specific to the Brazilian Chamber of Deputies, which they treat as \textsc{Against} \cite{Levorato2017}.

Some authors do not want to deal with weighted networks, generally because the analysis method they apply cannot handle them. It is quite straightforward to turn a weighted network into an unweighted one. However, these are similarity networks, which means they are \textit{fully connected}. This makes the unweighted networks difficult to analyze directly: authors need to apply an additional processing step to make them sparser. Crane \& Dempsey keep only the links whose weights are above a predefined threshold \cite{Crane2016}. For signed networks, it is possible to consider the absolute value of the weights \cite{Figueiredo2014, West2014, Levorato2017a}. In addition to this, Figueiredo \& Frota create both positive and negative links between the same two nodes when their agreement and disagreement levels are both above some threshold \cite{Figueiredo2014}. These filtering steps are also used by authors who just want to make their network sparser (without the methodological need for unweighted links) \cite{Balasubramanyan2010, Traag2013}. It is worth noticing that such simplification can be suspected of causing some information loss, though. One should therefore be cautious with this type of procedure and make sure that it preserves the important aspects of the considered system. Arinik \textit{et al}. study the effect of such filtering in terms of topological properties of the extracted networks, and performance of partitioning algorithms \cite{Arinik2017}. Based on their extensive computational experiments, they show empirically that, at least in their case, it eases the interpretation of the networks and of their analysis. These results are hardly generalizable to other situations, as this requires carefully exploring the space of possible threshold values~\cite{Figueiredo2014, Crane2016}.

In certain works, one objective is to study the \textit{polarization} of the considered institution, i.e. the way it splits in two or more antagonistic groups that constantly oppose each other over a significant period. For this purpose, they filter roll-calls in order to focus only on non-unanimous situations. Waugh \textit{et al}. discard roll-calls resulting in a minority group of less than $3\%$ of the legislators \cite{Waugh2011}. Macon \textit{et al}. only remove strictly unanimous roll-calls \cite{Macon2012}. Santamaria \& Gomez do not filter unanimous roll-calls, but they give less importance to situations of high general agreement through an entropy-based weight normalization \cite{Santamaria2015}.

Most authors analyze their vote similarity network using some form of partitioning method, in order to identify cohesive groups of voters. Certain works focus on the description of a new methodological tool, in which case the network is used as a benchmark to assess this method: community detection \cite{Balasubramanyan2010, Santamaria2015, Crane2016}, significance measure for community structures \cite{Traag2013}, visualization of dynamic community structures \cite{Reda2011}, resolution of the Maximum Balanced Subgraph problem \cite{Figueiredo2014}. Certain authors extract both signed and unsigned versions from the same dataset, and assess their informativeness by comparing the estimated partitions \cite{Macon2012, Mendonca2015}. Others  compare the partitions obtained by solving different partitioning problems on signed networks: Correlation Clustering (CC) and its relaxed version (RCC) \cite{Arinik2017}, CC and Symmetric Relaxed CC (SRCC) \cite{Levorato2017, Levorato2017a}. West \textit{et al}. use the vote similarity network as a part of their opinion prediction model, which also takes advantage of the content of the legislators' speeches \cite{West2014}. Other works are more driven by the political aspects and their applications. Waught \textit{et al}. use Newman \& Girvan's modularity~\cite{Newman2004e} to assess the quality of partitions. They interpret the modularity value as measuring the intensity of parliamentary polarization, which allows detecting important events, such as majority-party switches, or periods of (in)stability \cite{Waugh2011}. Dal Maso \textit{et al}. study the individual positions of the legislators through a perturbation-based approach \cite{Maso2014}. Levorato \& Frota study how the political crisis in Brazil affected the evolution of the voting groups at the Chamber of Deputies \cite{Levorato2017}. Levorato \textit{et al}. take advantage of their relaxed partitioning problem to identify groups of mediator countries at the UNGA \cite{Levorato2017a}.

The last point concerns the way time is managed. As mentioned before, a vote similarity network is processed by aggregating roll-calls over time. However, when the considered period is particularly long, or dense in terms of voting activity, it is possible to split it in several distinct periods, each one leading to a separated network. In most works, these are considered in a completely independent way \cite{Waugh2011, Macon2012, Traag2013, Maso2014, Figueiredo2014, Mendonca2015, Levorato2017, Levorato2017a, Arinik2017}, for instance by partitioning separately each network and then studying the evolution of the partition or of some measure of interest. But it is also possible to consider these graphs as time slices constituting a sequence, and therefore a dynamic network. This allows explicitly taking into account the coupling between successive time slices during the analysis. A few authors apply this principle to detect dynamic communities \cite{Reda2011, Mucha2010, Mucha2010a, Crane2016}. As explained for multiplex networks in the introduction, by comparison to a separate partition of each time slice, such a community can span \textit{several} graphs (time slices).

\subsection{Voter-Bill Networks}
\label{sec:RelWorkVoterBillNet}
All the works presented in the previous subsection induce two forms of information loss. First, they are based on a temporal integration of the raw data, performed when computing the similarity measure. In the case of unsigned variants, for instance, a weight of $0.5$ could mean that the considered legislators half agree all the time, or that they agree for half the roll-calls and disagree for the other half, or some combination of these situations. The problem also exist with signed variants, for instance a value close to $0$ can correspond to a situation where two legislators almost never participate in the same roll-calls, but also to a case where they agree half of the time. The fact that the vote similarity networks are unipartite projections of bipartite networks, which we call \textit{voter-bill networks}, is the second cause of information loss \cite{Latapy2008a}. 

A voter-bill network contains two kinds of nodes, representing legislators and bills. The links are signed unweighted, and represent the fact that a legislator voted \textsc{For} ($+$) or \textsc{Against} ($-$) some bill. Compared to the vote similarity networks, they allow avoiding both types of information loss, since, besides their bipartite nature, they require no temporal aggregation to separately represent all roll calls. However, maybe because the tools allowing to analyze bipartite graphs are less developed than their unipartite counterparts, only a few works use them as a modeling tool of voting behavior.

With vote similarity networks, the literature contains a number of different ways to handle absences and abstentions. However, this is not true for voter-bill networks: all authors represent these situations simply by an absence of link between the concerned voter and bill. Like for vote similarity networks, certain authors perform a filtering of the raw data, because they want to study how the considered system gets polarized. Some discard the strictly unanimous cases \cite{Mrvar2009, Macon2012}; Doreian \textit{et al}. \cite{Doreian2013} only focus on the supposedly more polarized military resolutions of the UNGA. Guimera \& Sales-Pardo work on a selection of the cases considered as simple (in particular, this involves no \textsc{Abstain} vote) \cite{Guimera2011}. Unlike for vote similarity networks, filtering roll-calls affects the problem size, by decreasing the number of nodes.

Like for vote similarity networks, most authors analyze this kind of networks through some node partitioning method. However, it is important to stress that due to their bipartite nature, the obtained clusters contain both types of nodes (voters and bills). For all the approaches relying on the structural balance concept, this means that a cluster contains voters tending to support and oppose the same bills, as well as the concerned bills. In \cite{Mrvar2009}, Mrvar \& Doreian use their blockmodeling approach to study the  US Supreme Court and detect different types of cases, for instance those causing a fundamental ideological divide between conservatives and liberals, or those resulting in a general consensus but disagreement on both extremes. Doreian \textit{et al}. apply a similar method on a larger dataset when analyzing the UNGA data \cite{Doreian2013}. In \cite{Guimera2011}, Guimera \& Sales-Pardo also study the US Supreme Court. They use their own stochastic blockmodeling approach to partition the network, and take advantage of this model to predict the decision of a justice given the behavior of the other justices for the same case. In \cite{Macon2012}, Macon \textit{et al}. use a modularity-based approach to study the UNGA voting activity, and remark that the evolution of this measure (in time) matches certain important diplomatic events. 

Although voter-bill networks  represent the whole dataset without information loss, doing so can lead to difficulties when interpreting their partitioning in case of large datasets. For this reason, authors generally split their dataset in several parts and extract a distinct network for each of them. In most situations, the authors use the temporal dimension: Macon \textit{et al}. extract one network for each annual UNGA session \cite{Macon2012}; Doreian \textit{et al}. divide the considered period in $4$ separate subperiods \cite{Doreian2013}. But this split can also be driven by the nature of the votes. For instance, when working on the Supreme Court, Mrvar \& Doreian remark that opinion prevalence can change widely from one case to the other, which makes it difficult to find a clear partition of both justices and cases \cite{Mrvar2009}. For this reason, they extract each network using only cases resulting in partitions whose clusters have predefined sizes, e.g. only 5 vs. 4 cases, or only 8 vs. 1 cases (there are 9 justices), and treat them separately. This is an important point for us, because it supports our own opinion, expressed in the introduction, that it is important to identify what Mrvar \& Doreian call the \textit{systematic patterns of voting behavior}. The method we propose can be considered as a way of automating this process, but there is also a more fundamental difference: we do not want to jointly partition voters and bills, but rather to partition the bills according to the voter partitions they induce.

In \cite{Macon2012}, Macon \textit{et al}. extract unsigned and signed vote similarity networks, as well as voter-bill networks from the same dataset, and compare them in terms of how they get partitioned by their method. They focus on certain specific sessions and explore a range of granularities for each of the $3$ network types. They find that the results are globally similar independently from the network type, even if bipartite ones theoretically convey more information. The main difference is that the clusters identified in the bipartite networks are generally larger. The authors highlight the complementary nature of the considered approaches.

Finally, the work presented by Wang \textit{et al}. in \cite{Wang2012aa} is an enhanced version of vote-bill networks. They extract what they call a \textit{heterogeneous network} from the US House of Representatives voting activity, for the 2008-10 period. Both legislators and bills are represented as nodes, and the network contains four types of links: political similarity between two legislators (based on co-sponsorship, party, age...), semantic similarity between two bills (based on their content), and support/opposition of a legislator to a bill (i.e. he voted \textsc{For}/\textsc{Against}). This amount to a vote-bill network with two additional types of similarity links connecting, on the one side, two legislators, and on the other side, two bills. They propose a model to predict the votes of legislators based on a custom random-walk taking place on the extracted network. 



\section{Data and Network Extraction}
\label{sec:Dataset}
Our goal is to compare our proposed method with the standard approach traditionally used when studying vote networks. We decided to use the results presented by Arinik \textit{et al}. \cite{Arinik2017} as a reference. Indeed, we expect our method to be able to answer some of the questions left open by their analysis. Moreover, both their dataset\footnote{\url{https://doi.org/10.6084/m9.figshare.5785833}} and source code\footnote{\url{https://github.com/CompNet/NetVotes}} are publicly available online. In this section, we present the dataset and highlight its most relevant characteristics (Section~\ref{sec:DataDescr}). We then summarize how the signed graph constituting the layers of our multiplex signed network are extracted from the raw voting data (Section~\ref{sec:DataExtr}).

\subsection{IYP Dataset}
\label{sec:DataDescr}
The raw data are publicly available on the official website of the EP\footnote{\url{http://www.europarl.europa.eu/}}, but they are technically difficult to collect. Some citizen oversight websites did the work of retrieving them, and publishing them online. Arinik \textit{et al}. have used such a dataset, provided by the website \textit{It's Your Parliament}\footnote{\url{http://www.itsyourparliament.eu/}} (IYP).

These data describe the activity of the MEPs during the 7\textsuperscript{th} term of the EP (2009--14). They are constituted of the votes cast by all MEPs for all roll-calls taking place during a plenary at the EP. Note that the default voting procedure at the EP is the show of hands, during which individual votes are not recorded, and that roll-calls happen only under certain circumstances. During the 7th term, these noticeably include the final vote of any legislative proposition, or a written demand by a group of $40$ MEPs or a parliamentary group~\cite{EP:Rules}. These data are therefore incomplete by definition, but recent studies have shown they are nevertheless representative of the overall voting behavior \cite{Kaniok2017}. As mentioned in Section~\ref{sec:RelatedWork}, several works have studied them under the form of networks \cite{Traag2013,Mendonca2015,Santamaria2015,Arinik2017}.

The vote of a MEP can take $3$ distinct values: \textsc{For} (the MEP supports the proposition), \textsc{Against} (he opposes the proposition) and \textsc{Abstention} (he does not want to take a stand). It is also possible for the MEP not to vote at all. Officially, the EP distinguishes various types of absences or reasons for not voting (see \cite{Mendonca2015}). But in this dataset, all are simply represented by the value \textsc{Absent}. Each text is itself associated to one among $21$ specific policy domains (see \cite{Mendonca2015} for the complete list).

A number of personal details are available for each MEP: 
\begin{itemize}
	\item \textit{Name}: the full name of the MEP;
    \item \textit{Country}: one of the (then) $28$ member states in which the MEP was elected; 
    \item \textit{Party}: the national political party to which the MEP belongs, in his own country;
    \item \textit{Group}: the parliamentary group to which the MEP belongs, at the EP.
\end{itemize}

The groups are important when interpreting a partition of the MEP set, because they correspond to the political position that MEPs are supposed to hold, at least theoretically. During the 7\textsuperscript{th} term, they were (in decreasing order of size): 
\begin{itemize}
	\item \textit{European People's Party} (EPP): right/center-right conservatives;
    \item \textit{Progressive Alliance of Socialists and Democrats} (S\&D): center-left;
    \item \textit{Alliance of Liberals and Democrats for Europe} (ALDE): right/center-right neoliberals;
    \item \textit{Greens--European Free Alliance} (G-EFA): left environmentalists, progressives and regionalists;
    \item \textit{European Conservatives and Reformists} (ECR): right euroskeptics and anti-federalists;
    \item \textit{European United Left--Nordic Green Left} (GUE-NGL): left/far-left, socialists and communists;
    \item \textit{Europe of Freedom and Democracy} (EFD): right/far-right euroskeptics;
    \item \textit{Non-Inscrits} (NI, non-attached members): technical group containing members not belonging to any of the other groups (during this term: mainly far-right MEPs).
\end{itemize}
One would expect that partitions automatically estimated based on the voting data would fit this division, but as we will see in Section~\ref{sec:Results}, this is not necessarily the case.

Arinik \textit{et al}. worked on the whole dataset, but they also focused on some subsets, in order to deal with smaller data and perform a more thorough and qualitative interpretation of their results. In particular, they restrained some of their analysis to the French and Italian MEPs, to propositions related to the \textit{AGRI} (agriculture) policy domain, and considered separately the years constituting the term. This domain was selected due to its potentially polarizing nature: the \textit{Common Agricultural Policy} (CAP, cf. Appendices~\ref{sec:appendix-definitions} and \ref{sec:appendix-reforms} for more context) has historically been of great importance for Europe, especially for France~\cite{CapOverviewFr} and Italy~\cite{CapOverviewIt}, due to the importance of agriculture in their economies (38\% of the total EU budget~\cite{EPCouncilCAP}), but a part of the population wants to leave the industrial model of production. Also, 2013 CAP reforms cover various aspects in agriculture (e.g. production quotas, environmental aspects, market regulation, etc.), this would allow to analyze the results in multiple perspectives. For these reasons, and in order to compare our own results to theirs, we work with the same subset of the original dataset. 

At some point of our processing, we will need to characterize subgroups of legislative propositions in a topical way, in order to assess their thematic homogeneity (or lack thereof). Since all the propositions we examine are related to agriculture, we have to consider subdomains. For this purpose, we use the typology proposed by EUR-Lex, the EU website for the publication of official documents such as treaties and legislation. The detailed hierarchy of domains is described in Appendix~\ref{sec:appendix-DataCat}. Appendix~\ref{sec:appendix-Positions} details the position of each EP group on the main AGRI-related topics for the considered 2012-13 period.

\subsection{Network Extraction}
\label{sec:DataExtr}
The voting behavior of each MEP is represented by a series of vote values, each one corresponding to a specific roll-call of the considered period. It is therefore possible to use both paradigms presented in Section~\ref{sec:RelatedWork}. We decide not to use a voter-bill network, because its bipartite nature is not compatible with our objective of identifying the characteristic voting behavior patterns of the EP. As mentioned in Section~\ref{sec:RelWorkVoterBillNet}, partitioning such a network would amount to jointly partitioning the MEPs and propositions, whereas we want to partition the propositions \textit{according to} the way the MEPs vote. 

We therefore select the family of the vote similarity networks. However, as mentioned in the introduction, we want to avoid the information loss provoked when averaging the vote similarity values over the considered series of votes. A natural approach consists in modeling the data as a dynamic network, in which each time slice corresponds to a roll-call. However, this is not appropriate for the EP, due to its very specific scheduling, which differs much from a typical national parliament's. At the EP, there are only $4$ days of plenary session by month, in average, during which the MEPs consider a large number of propositions. The chronology of the propositions therefore has little effect on the votes, as we verified empirically. Instead, we turn to a multiplex representation, in which each layer models one roll-call.

Let us now introduce our notations to handle graphs. Let $G=(V,E)$ be an \textit{undirected graph}, where $V$ and $E$ are the sets of vertices and edges, respectively. We note $n=|V|$ and $m=|E|$ the numbers of vertices and edges, respectively. Consider a function $s : E \rightarrow \{+,- \}$ that assigns a \textit{sign} to each edge in $E$. An undirected graph $G$ together with a function $s$ is called a \textit{signed graph}, denoted by $G = (V, E, s)$. An edge $e \in E$ is called negative if $s(e) = -$ and positive if $s(e) = +$. We note $E^-$ and $E^+$ the sets of negative and positive edges in a signed graph, respectively. We note $G = \{ G_1,...,G_p \}$ a \textit{multiplex signed graph} constituted of $p$ layers. Each layer $i\in\{1,\ldots,p\}$ is itself a signed graph $G_i = (V,E_i,s_i)$. Note that all graphs $G_i$ have the same node set $V$, but possibly different structures.

The similarity function that we use is very basic, since it is applied to each roll-call considered separately (by opposition to the whole series). For a pair of MEPs $u$ and $v$ and a roll-call $i$, we have $(u,v) \in E_i$ if and only if both MEPs voted, i.e. neither were absent. Moreover, we set $s_i(u,v)=+$ if both MEPs voted similarly (\textsc{For}-\textsc{For}, \textsc{Against}-\textsc{Against}, or \textsc{Abstain}-\textsc{Abstain}) and $-$ otherwise. Unlike other approaches such as \cite{Arinik2017}, we do not need to filter the resulting links, because their weights are not the result of some averaging, and all of them are assumed to be informative. As described in Section~\ref{sec:RelWorkVoteSimNet}, it is difficult to decide how to treat abstention, since it can be considered as the expression of some intermediate position. In this extraction phase, we consider \textsc{Abstain} exactly like the other forms of vote (\textsc{For} and \textsc{Against}). However, we later propose and compare two different ways of handling abstention in the rest of the process.

\section{Methods}
\label{sec:Methods}
In this section, we describe the method that we propose to analyze our multiplex signed networks, whose extraction process is presented in Section~\ref{sec:DataExtr}. We will be handling various types of partitions, so we need to clarify our terminology first. We call \textit{voting behavior pattern} of the EP (or \textit{pattern}, for short) a partition of the set of all MEPs obtained for a given roll-call (e.g. GUE-NGL and Greens vs. the rest). The subsets of MEPs constituting this partition are called \textit{factions} (e.g. a coalition of GUE-NGL and Greens). The proposition considered at a given roll-call (bill, amendment, motion...) is generically called a \textit{document}. A pattern therefore represents the way the parliament is split at the occasion of a roll-call concerning a specific document. We reserve the term \textit{clustering} to refer to a partition of the set of all patterns (i.e. the patterns corresponding to all considered documents). Since each pattern describes the EP behavior for a given document, a clustering can also be interpreted as a partition of the set of all documents. The subsets of documents (or patterns) constituting a clustering are simply called \textit{clusters}. For instance, if documents \#1, 2 and 3 form a cluster, this means the pattern was similar for all three corresponding roll-calls.

\begin{figure*}[!htb]
	\centering
    \includegraphics[width=\textwidth]{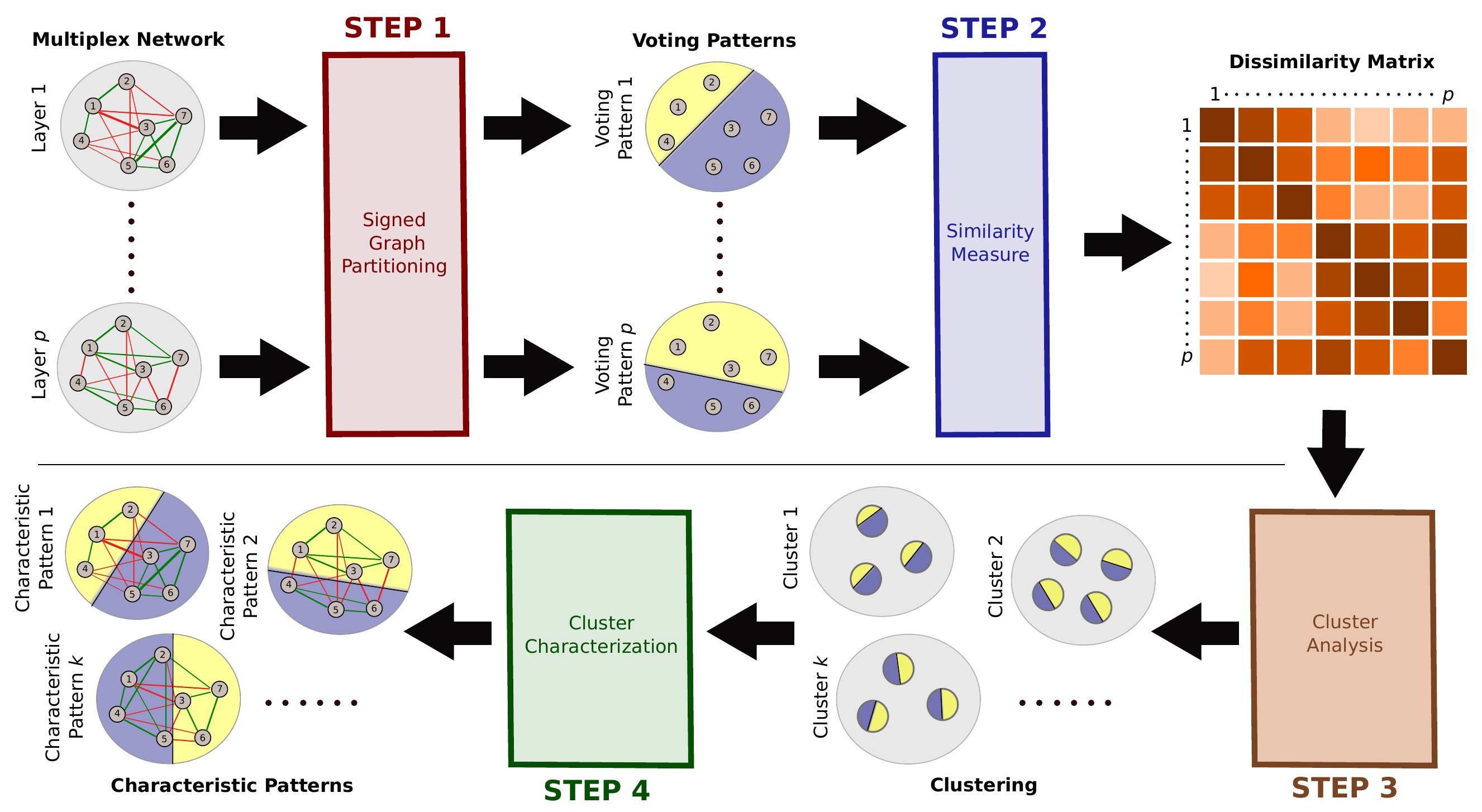}
	\caption{General workflow of the proposed analysis method.}
	\label{fig:workflow}
\end{figure*}

The goal of our method is to identify the main types of patterns occurring at the EP, and to characterize them in terms of political groups, individual MEPs and topics of the concerned documents. To this aim, we propose a four-stepped method, summarized in Figure~\ref{fig:workflow}, and detailed in the rest of this section. The first step is to separately partition the $p$ layers of our multiplex signed network, in order to get as many patterns (Section~\ref{sec:ProcessingPatterns}). Unlike some approaches mentioned in Section~\ref{sec:RelatedWork}, which need to filter out unanimous roll-calls, our method automatically treat unanimity as just another type of pattern. The second step consists in computing the similarity between the patterns (Section~\ref{sec:ComputingSimilarity}), in order to perform the third step, which is a cluster analysis (Section~\ref{sec:PerformingClustering}). This leads to a set of clusters, each one gathering similar patterns. The fourth step is to process what we call the \textit{characteristic pattern} of each cluster, which is supposed to consensually represent all the patterns belonging to the cluster (Section~\ref{sec:CharacteristicPatterns}). This results in a set of characteristic patterns, each one associated to a cluster of documents, which can be used as the basis of the interpretation work.

\subsection{Processing the Patterns}
\label{sec:ProcessingPatterns}
Detecting the pattern associated to the $i^{th}$ document amounts to partitioning the $i^{th}$ layer of our multiplex signed network, which represents the $i^{th}$ roll-call of the considered collection. This layer is an unweighted signed graph $G_i$ whose nodes are connected depending on how the MEPs they represent voted during this roll-call. MEPs who voted similarly are connected together by positive links, and are connected by negative links to MEPs that voted differently from them. MEPs who did not vote at all (\textsc{Absent}) are isolates (nodes without any neighbor). 

We consider that absent MEPs should not affect the pattern, since they did not express any opinion regarding the matter at hand during this roll-call, so we simply ignore them at this stage. For those who abstained, we consider two options. The first assumes they are opinionless too, and therefore also removes them from the graph. The second supposes (as other authors do, cf. Section~\ref{sec:RelWorkVoteSimNet}) that abstaining is a way of expressing some middle ground position, and keeps them in the graph. When taking the rest of the process into account, it turns out the second option is more informative as it can detect factions of abstentionists. For this reason, we only discuss this second option in the rest of the document.

Next, we need to identify the factions of similarly voting MEPs remaining in the graph, which can be done by solving the \textit{Correlation Clustering} problem (CC). In its original version \cite{Bansal2002}, and consistently with the definition of structural balance given earlier in the introduction, it consists in finding a partition of the set of vertices which maximizes both the number of positive links located \textit{inside} the subgroups, and that of negative links located \textit{between} them. In order to identify this partition, we use Ex-CC (Exact Correlation Clustering), which is an integer programming-based method able to solve the CC problem exactly \cite{Ales2016}.

\subsection{Computing the Similarity Values}
\label{sec:ComputingSimilarity}
At this stage, we have identified the pattern associated to each roll-call. We now want to gather similar patterns together. For this purpose, we use a classic cluster analysis approach.  

We must first process the dissimilarity matrix by comparing each pair of patterns. A number of measures have been defined to compare such partitions, each one possessing a specific behavior. We considered the $4$ most widespread such measures \cite{Labatut2014}: the Purity \cite{Manning2008}, the Rand Index (RI) \cite{Rand1971} and its adjusted version (ARI) \cite{Hubert1985}, and the Normalized Mutual Information (NMI) \cite{Strehl2002}. The \textit{Purity} works by assessing the maximal overlap between the clusters of the considered partitions. Unlike the other mentioned measures, it is asymmetric, i.e. its value can change if we switch the partitions. For this reason, it is customary to use its harmonic mean. The \textit{Rand Index} is based on pairwise comparisons: it corresponds to the proportion of pairs of elements treated similarly in both compared partitions (i.e. placed in the same clusters, or in two different clusters). The \textit{Adjusted Rand Index} is a chance-corrected version of the Rand Index, i.e. it accounts for chance agreement between the partitions. Finally, the \textit{Normalized Mutual Information} considers both partitions as discrete random variables and assesses their similarity through their probabilistic dependence (mutual information). The normalization allows getting a fixed upper bound, like the other mentioned measures.

These measures are sensitive to various characteristics of the compared partitions, such as the sizes or numbers of the clusters. We want the selected measure to be sensitive enough so that it detects relevant cluster changes, but not enough so that it is affected by simple individual exchanges of MEPs. We proceed by enumeration to express this preference. First, we manually design a set of archetypical partitions, i.e. partitions representing specific situations of interest (see the example right after). Second, we order pairs of these partitions in terms of how much different we consider them to be. Third, we compute the $4$ previously mentioned measures for these pairs, and select the measure matching best our ranking. Here is an example of a pair of archetypical partitions. Suppose we have two partitions of $20$ MEPs: each one has a $19$ MEPs faction and a $1$ MEP one, but the latter is different in both partitions. Intuitively, we consider these partitions as very similar. The NMI, however, returns a value close to zero, when the other measures consider these partitions as rather similar. Over the set of considered partitions, the Purity and the Rand Index both clearly emerge as the most appropriate measures given our constraints and needs, with a slight advantage to the Purity.

Based on the Purity, we compute a similarity value for each pair of patterns. Note that when a MEP is present in one pattern but not the other, we just ignore him. We then build a dissimilarity matrix summarizing these comparisons.

\subsection{Performing the Clustering}
\label{sec:PerformingClustering}
Next, we apply the $k$-medoids clustering method to the previously obtained similarity matrix \cite{Kaufman2009}. It is similar to the well-known $k$-means algorithm in the sense that it tries to partition the dataset in $k$ clusters, while minimizing the distance between the members of each cluster and some center of the cluster. The difference is that in $k$-means, this center is an average value, whereas in $k$-medoids it is one of the actual data point from the dataset. It is generally used in place of $k$-means when one cannot perform the required average operation, which is our case (we cannot straightforwardly process an average pattern). 

This method requires us to specify $k$, but we do not know it in advance. In this situation, the standard approach is to use all possible values of $k$, from $2$ to $n$ (where $n$ is the number of patterns), and assess the quality of the $n-1$ resulting clusterings through some internal criterion. The most widespread such measure is the \textit{Silhouette} $S$, which characterizes the clustering in terms of internal cohesion and external separation of the clusters \cite{Rousseeuw1987}. It takes a value between $-1$ and $+1$, where the latter represent the best possible clustering. 

In theory, the $k$ value associated to the highest Silhouette is the best candidate. However, in practice, one possibly has to consider other factors to make his choice. For example, marginal improvements of the Silhouette are sometimes caused by the creation of singleton clusters, which generally do not bring much relevant information in terms of interpretation of the clustering. It is therefore necessary to study qualitatively how the clusters evolve with $k$ to make an informed choice. Some constraints can also arise from the application context itself. In our case, each cluster corresponds to a typical way of 
classifying
the MEPs. Yet, there are only a small number of political groups at the EP. Moreover, it is very unlikely that the voting activity displays all mathematically possible combinations of these groups, for matters of ideological incompatibility. Therefore, we expect an informative clustering to contain a small number of clusters, say much fewer than $10$. Note that larger values of $k$ are not uninteresting though, but they are likely to correspond to finer differences between patterns, e.g. concerning one or a few MEPs switching sides.

To be exhaustive, we have repeated the same clustering process using the $3$ other partition similarity measures mentioned earlier instead of the Purity. It turns out the general quality of the obtained clusterings is much higher for Purity ($\approx 0.65$) and RI ($\approx 0.67$) than for NMI ($\approx 0.36$) and ARI ($\approx 0.44$), which confirms the assessment performed before on manually designed partitions.

\subsection{Computing the Characteristic Patterns}
\label{sec:CharacteristicPatterns}
We now have $k$ clusters, each one containing a certain number of patterns. The patterns constituting a cluster may differ slightly, but overall they are supposed to be very similar. The next step is to compute a characteristic pattern representing the whole cluster, such that these small differences are smoothed. 

For this purpose, we use a similarity network-based approach, inspired by the work of Lancichinetti \& Fortunato \cite{Lancichinetti2012}. Based on a collection of $n$ partitions of the same set, they derive a consensual partition by first extracting a weighted similarity network, and then performing community detection in this network. The resulting communities correspond to consensual clusters, and the community structure is the consensual partition. Their network is built as follow: each node represent an element of the partitioned set, and the weight of link is the proportion of partitions in which both connected nodes belong to the same cluster. We experimented with this approach, and found out we obtain better results (at least for our data) by using the following \textit{signed} version. The weights are now the difference between the proportion of patterns putting both MEPs in the same cluster, and the proportion of patterns putting them in different ones. Moreover, when at least one MEP is absent from a pattern, it is not taken into account when computing the proportions. Finally, we use Ex-CC instead of a community detection algorithm, to identify a partition corresponding to the characteristic pattern of the cluster. 

Like any pattern, a characteristic pattern can take one of three forms: a single faction in case of unanimity, $2$ antagonistic factions if some MEPs supported the concerned documents whereas others opposed them, and $3$ in case of an additional faction of abstentionists. Our method is therefore able to handle (quasi-)unanimity as just a specific case of pattern, and does not require any preprocessing related to this point, unlike some approaches mentioned in Section~\ref{sec:RelatedWork}.

As discussed in Section~\ref{sec:RelatedWork}, the management of absences is a delicate point. In our case, as for other authors adopting a similar approach, excluding the absences when processing the weights of our similarity network can give much importance to the few votes cast by a chronically absent MEP. We solve this issue by filtering the similarity network \textit{before} applying Ex-CC: we remove all MEPs who did not participate in at least half the roll-calls constituting the cluster. In practice, the situation is quite rare, and this concerns only a few MEPs in a few clusters. One advantage, as we will see in Section~\ref{sec:Results}, is that it allows detecting MEPs whose absence coincides only with certain patterns (and therefore, certain subsets of legislative documents).

One could argue that what we do here when identifying the characteristic patterns, amounts to aggregating roll-calls, which is true in a way. Yet, we pointed out earlier that this leads to some information loss, which in turns make it difficult to interpret results such as the ones presented in \cite{Arinik2017}. However, this point holds only when considering the whole collection of roll-calls at once, because this results in mixing all the distinct voting behaviors exhibited by MEPs. In our case, we first detect clusters of similar pattern, which means that the roll-calls we aggregate are, by construction, very similar. Thus, this limitation does not apply.

\section{Results}
\label{sec:Results}
As explained before, we want to illustrate how our method differs from the classic vote similarity network approach, which relies on the temporal integration of the raw vote data. For this purpose, we analyze the same data as Arinik \textit{et al}. \cite{Arinik2017}, and more particularly the votes of French and Italian MEPs related to agricultural questions during the 2012-13 legislative year. We take their discussion as a reference when commenting our own results, highlighting both differences and similarities between these approaches. Our source code is publicly available online\footnote{https://github.com/CompNet/MultiNetVotes}.

As mentioned before, Appendices~\ref{sec:appendix-definitions}, \ref{sec:appendix-reforms} and \ref{sec:appendix-DataCat} offer more context regarding the CAP, and Appendix~\ref{sec:appendix-Positions} is a summary of the positions of the EP groups relatively to these topics. The rest of this section is organized as follows. We first introduce the conventions used to describe both French and Italian cases (Section~\ref{sec:ResultsGeneral}), before turning to the results themselves. One section is dedicated to the French MEPs (\ref{sec:ResultsFr}) and another to the Italian ones (\ref{sec:ResultsIt}).

\subsection{Conventions and General Remarks}
\label{sec:ResultsGeneral}
To denote the clusters, we use a notation indicating first the concerned country ($Fr$ for France and $It$ for Italy), then the value of $k$ used when detecting the considered clustering, and finally the number of the cluster in this clustering. For instance: $Fr$-$k3$-$clu2$ is the name of the second cluster of the clustering obtained with $k=3$ for the French MEPs.




We represent all networks and voting patterns using a circular layout (Figures~\ref{fig:fr-all-graphs} and \ref{fig:it-all-graphs}) generated through Circos\footnote{\url{http://circos.ca}}. We describe them generically here, for matters of convenience. They shall be read from the center to the periphery. The negative and positive links are drawn at the center, in red and green, respectively. Next, the inner colored ring represents the nodes (MEPs), and these colors correspond to the factions constituting the detected pattern (i.e. partition of the MEPs). If a MEP was often absent, he is ignored (as explained in Section~\ref{sec:CharacteristicPatterns}) and appears in white. The names of the MEPs are indicated separately in Appendix~\ref{sec:appendix-fr-k5-clu1} (Figure~\ref{fig:appendix-fr-it-graph-legend}). Finally, the outer ring shows the European political groups to which the MEPs belong. They are ordered according to the political spectrum, from left to right: GUE-NGL (red), G-EFA (green), S\&D (pink), ALDE (orange), EPP (light blue), ECR (dark blue), EFD (purple) and NI (brown) (see Section \ref{sec:DataDescr} for their full names and descriptions).

\subsection{Interpretation of a specific case: France}
\label{sec:ResultsFr}
We first focus on the results associated with the French MEPs. We summarize the findings of Arinik \textit{et al}. (\ref{sec:ResultsFrArinik}), then turn to our own clustering results (Section~\ref{sec:ResultsFrClustering}), and finally discuss the obtained characteristic patterns and compare them with Arinik \textit{et al}.'s (Section~\ref{sec:ResultsFrPatterns}).

\subsubsection{Baseline}
\label{sec:ResultsFrArinik}
Arinik \textit{et al}. \cite{Arinik2017} have applied a classic method, and extracted a vote similarity network similar to the ones described in Section~\ref{sec:RelWorkVoteSimNet}, by integrating their data over the whole considered legislative year (2012-13). Moreover, they have filtered the weakest links to sparsify the network. In the rest of this section, we call it the \textit{integrated network}, by opposition to the similarity networks that our method extracts \textit{for each distinct cluster}, based on which the characteristic patterns are detected (Step 4 of our method, Section~\ref{sec:CharacteristicPatterns}).

\begin{sidewaysfigure}
    \centering
    \begin{subfigure}[b]{6.8cm}
	    \centering
        \includegraphics[width=1.1\textwidth]{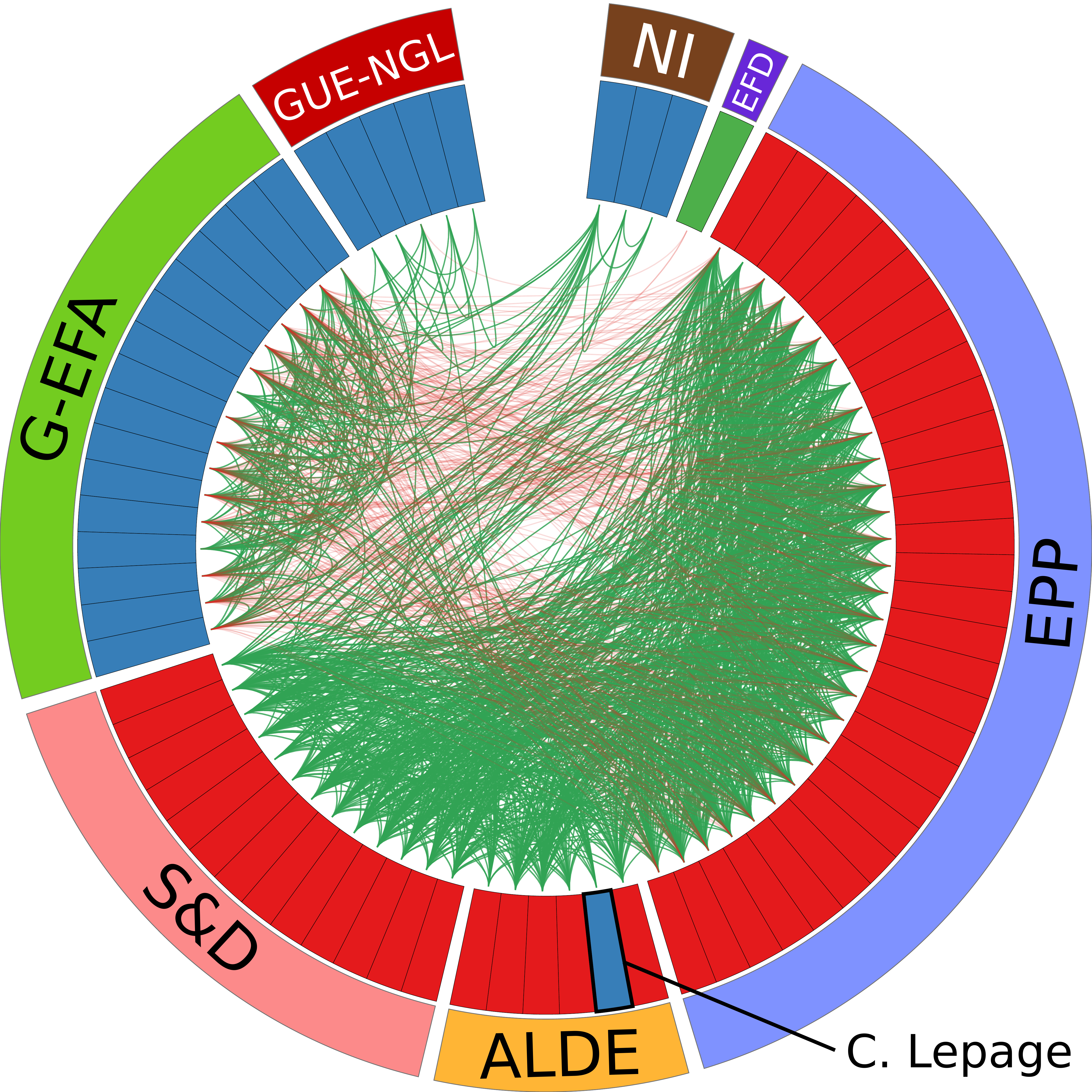}
        \caption{CC (Arinik \textit{et al}.)}
        \label{fig:CC-France-iknow}
    \end{subfigure}
    \hspace{1cm} 
    \vrule{}
    \hspace{0.05cm} 
    \begin{subfigure}[b]{6.8cm}
	    \centering
        \includegraphics[width=1.1\textwidth]{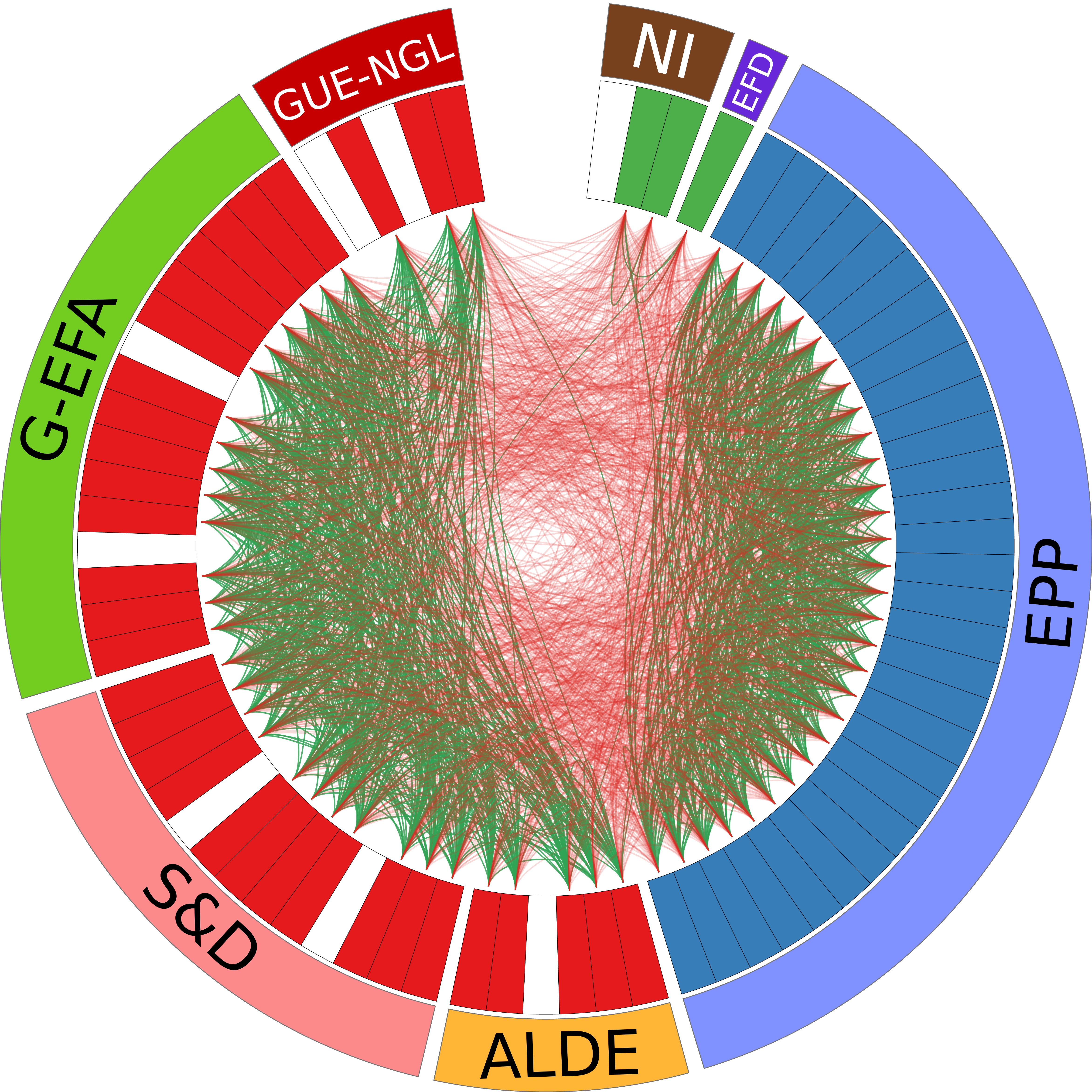} 
        \caption{$Fr$-$k5$-$clu2$}
        \label{fig:fr-k5-clu2}
    \end{subfigure}
    \hspace{1cm} 
    \begin{subfigure}[b]{6.8cm}
	    \centering
        \includegraphics[width=1.1\textwidth]{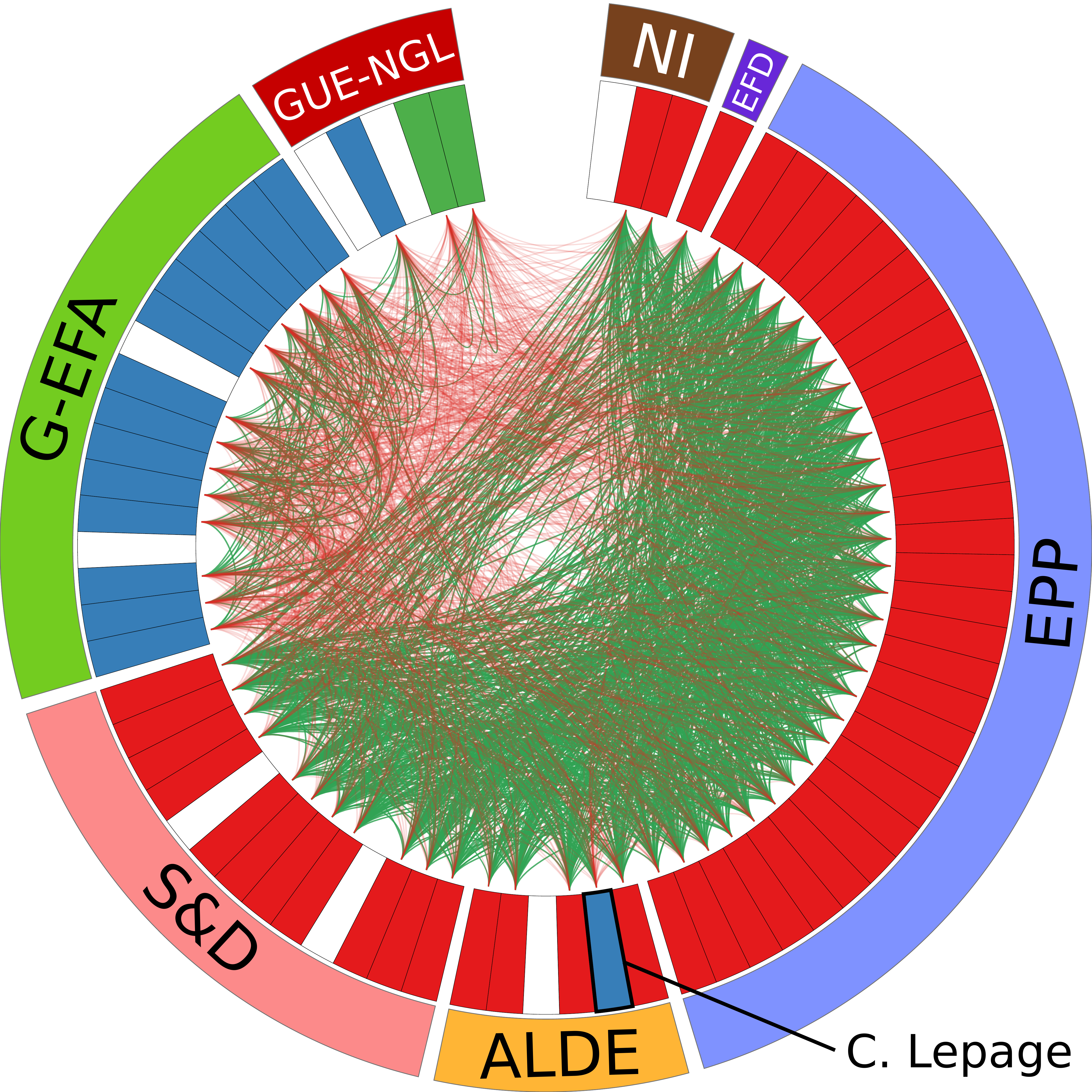}
        \caption{$Fr$-$k5$-$clu3$}
        \label{fig:fr-k5-clu3}
    \end{subfigure}
    \\
    \begin{subfigure}[b]{6.8cm}
	    \centering
        \includegraphics[width=1.1\textwidth]{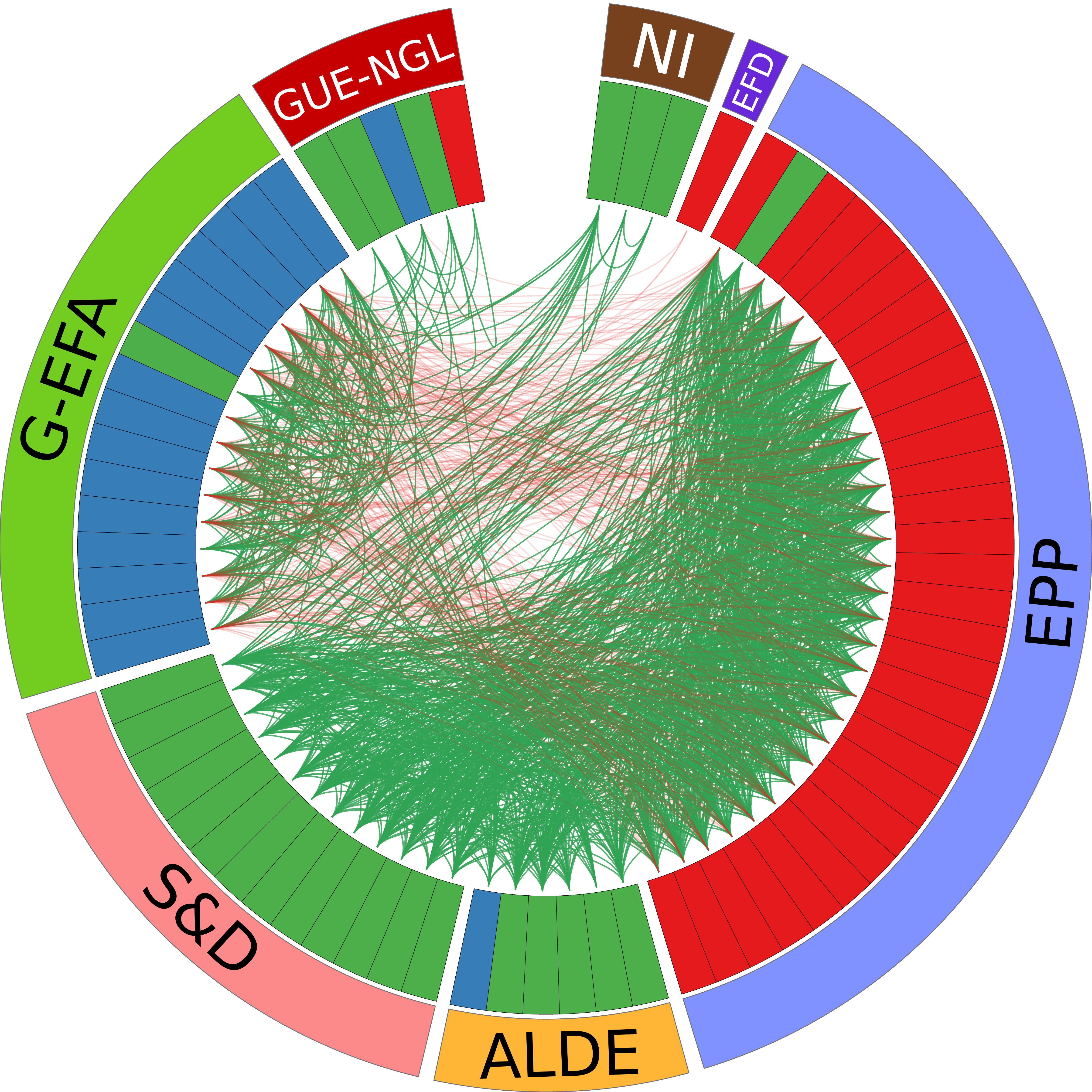}
        \caption{RCC (Arinik \textit{et al}.)}
        \label{fig:RCC-France-iknow}
    \end{subfigure}
    \hspace{1cm} 
    \vrule{}
    \hspace{0.05cm} 
    \begin{subfigure}[b]{6.8cm}
	    \centering
        \includegraphics[width=1.1\textwidth]{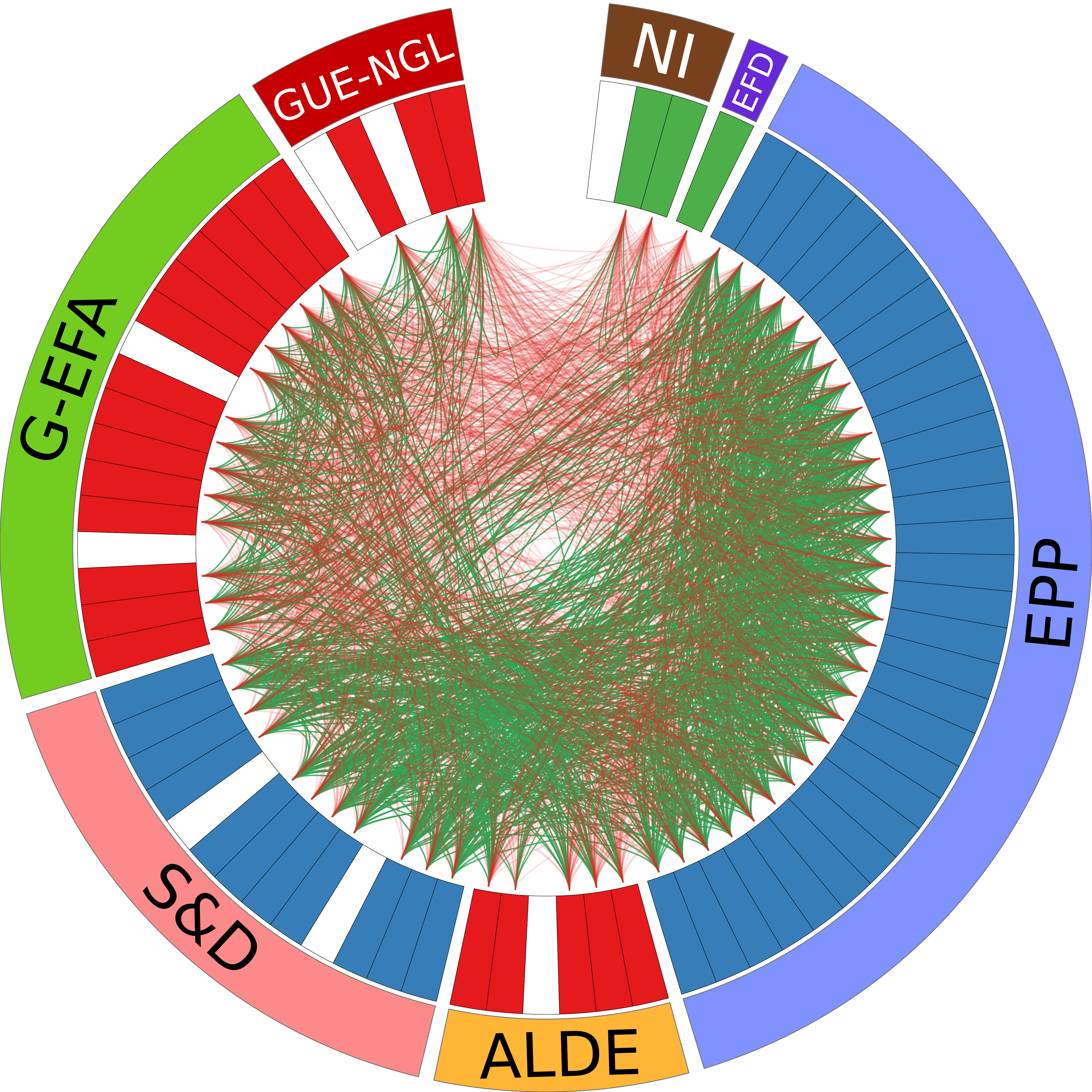}
        \caption{$Fr$-$k5$-$clu4$}
        \label{fig:fr-k5-clu4}
    \end{subfigure}
    \hspace{1cm} 
    \begin{subfigure}[b]{6.8cm}
	    \centering
        \includegraphics[width=1.1\textwidth]{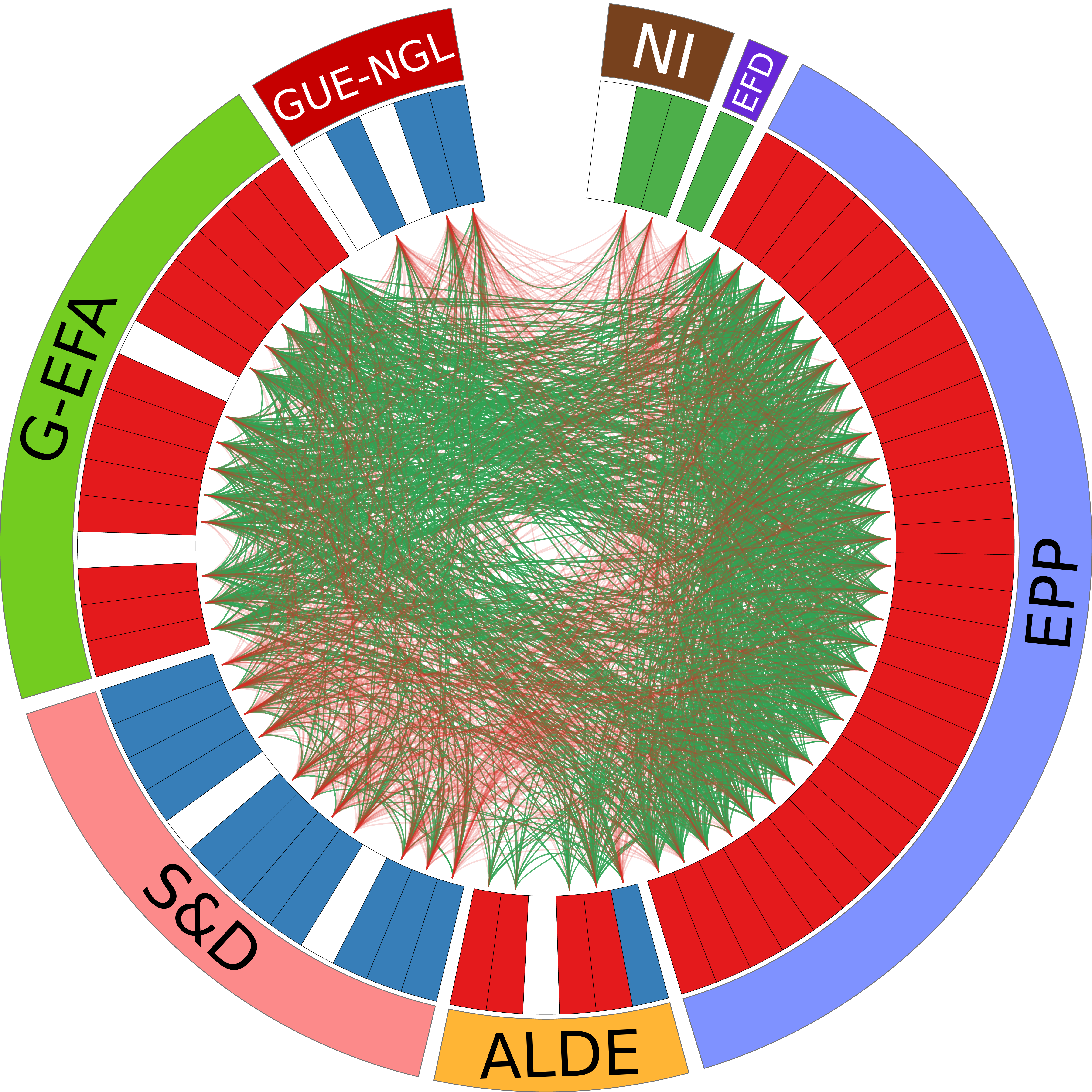}
        \caption{$Fr$-$k5$-$clu5$}
        \label{fig:fr-k5-clu5}
    \end{subfigure}
    \caption{Voting behavior patterns of the French MEPs on AGRI questions in 2012-13. Red and green lines at the center represent negative and positive links, respectively (red links are drawn on top of green ones in order to improve readability). Around the links, each MEP is represented by a colored tile, whose color corresponds to the MEP's faction in the displayed pattern. The green factions in plots (b), (c), (e) and (f) correspond to abstentionists. The MEPs' names are indicated separately in Figure~\ref{fig:appendix-fr-it-graph-legend}. The outer ring represents the political groups at the EP. The left plots show the patterns obtained by  Arinik \textit{et al}. \cite{Arinik2017} on the same integrated network when solving (a) CC and (d) RCC. The right plots show the 2nd to 5th clusters obtained with our proposed method for $k=5$: (b) $Fr$-$k5$-$clu2$ (\%15 of roll-calls), (c) $Fr$-$k5$-$clu3$ (\%32 of roll-calls), (e) $Fr$-$k5$-$clu4$ (\%8 of roll-calls), and (f) $Fr$-$k5$-$clu2$ (\%3 of roll-calls). The first cluster, $Fr$-$k5$-$clu1$, which corresponds to a unanimity situation, is represented separately in Figure~\ref{fig:appendix-fr-k5-clu1}.}
    \label{fig:fr-all-graphs}
\end{sidewaysfigure}

Figure~\ref{fig:CC-France-iknow} and Figure~\ref{fig:RCC-France-iknow} show the best partition that Arinik \textit{et al}. obtained by solving the CC and RCC problems, respectively. They state that the network is highly polarized, as it contains many negative links and results in good partitions in terms of structural balance. They identify two antagonistic factions respectively led by the environmentalists (G-EFA) and the right conservatives (EPP), joined by some other groups or individual MEPs. The position of S\&D and ALDE is interesting, because they belong to the right-wing faction according to CC, whereas they hold an intermediate position with RCC. Arinik \textit{et al}. were not able to give a solid explanation for this, or to discover in which context this happens. But they assumed that these groups were sometimes voting like the left-wing faction, and sometimes like the right-wing one.


\subsubsection{Clustering}
\label{sec:ResultsFrClustering}
\begin{figure}[!htb]
  \begin{tabular}{lr}
    \begin{minipage}{0.45\linewidth}
    	\includegraphics[width=1.0\linewidth]{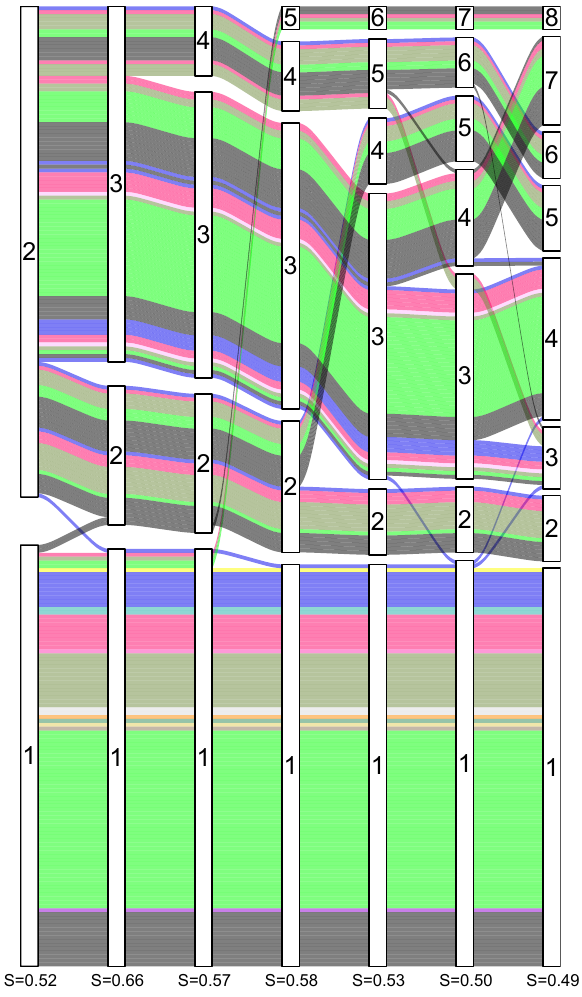}
    \end{minipage} 
    	& 
	\begin{minipage}{0.53\linewidth}
      \centering
      
      \begin{subfigure}[b]{1.0\textwidth}
      \centering
      \hspace{-0.40cm}
         \includegraphics[width=0.8\textwidth]{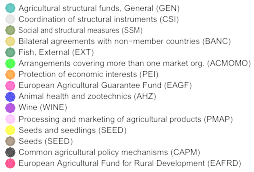}
      \end{subfigure}
      
		\begin{subfigure}[b]{1.0\textwidth}
        \hspace{-0.40cm}
          \includegraphics[width=1.0\textwidth]{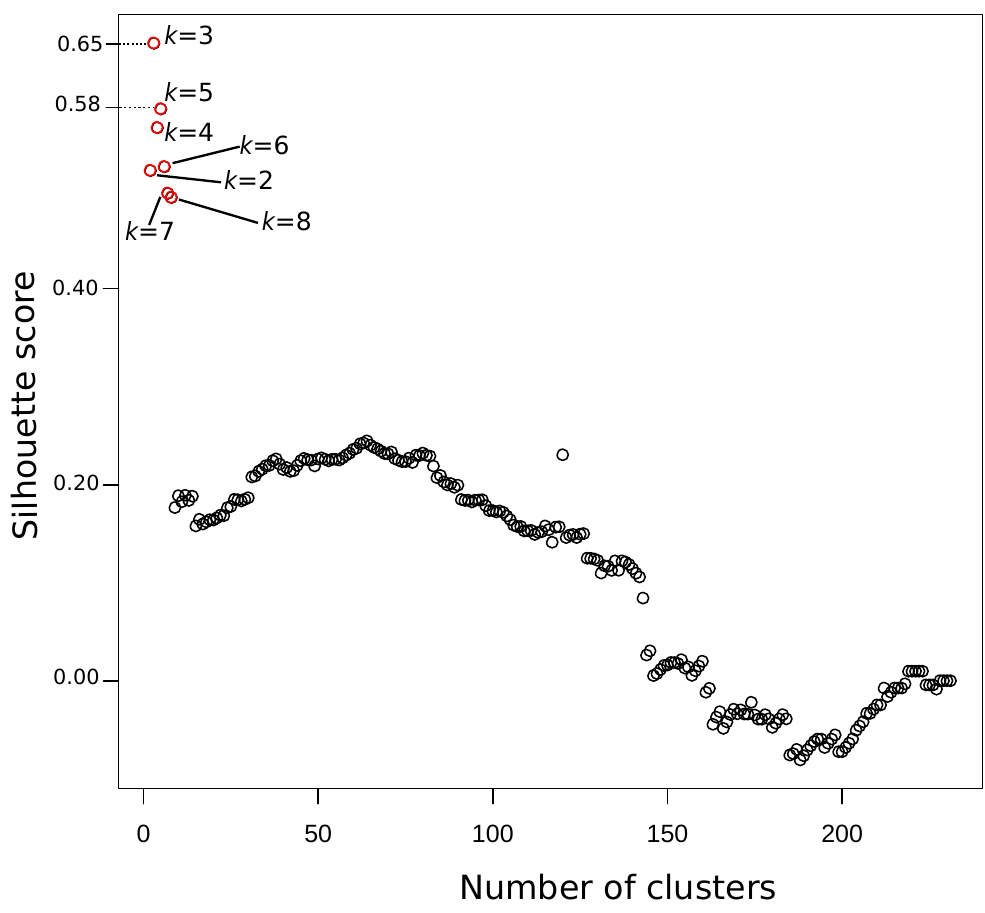}
      	\end{subfigure}
    \end{minipage}
  \end{tabular}
  \caption{Alluvial diagram (left) with its legend (top right), and Silhouette scores (bottom right) for the French MEPs.}
  \label{fig:fr-sil-alluvial}
\end{figure}

Figure~\ref{fig:fr-sil-alluvial} displays the results obtained after the third step of our method ($k$-medoids clustering, Section~\ref{sec:PerformingClustering}), when applied to the same raw data. The bottom right plot shows the Silhouette score as a function of the $k$ value used when performing the clustering. The highest Silhouette scores are obtained for the smallest $k$ values. Note, in particular, the very large gap between $k=1,...,8$ and $k \geq 9$. This confirms our assumption from Section~\ref{sec:PerformingClustering} regarding the expected number of clusters, and justifies that we discard the clusterings obtained for $k \geq 8$ in our analysis. 

The left plot of the same figure is an alluvial diagram\footnote{\url{https://cran.r-project.org/web/packages/alluvial/}} representing the changes underwent by the clustering depending on $k$. Each vertical bar corresponds to the clustering obtained for a given value of $k$, ranging from $2$ to $8$. Its numbered vertical rectangles represent the constituting clusters. The value indicated below the bar is the corresponding Silhouette score. Each horizontal line represents a document, and its color depends on the document subdomain (cf. Section~\ref{sec:appendix-DataCat}). If a document has several subdomains, it is duplicated as many times (but here, this concerns only $3$ documents).


Let us first discuss the topical heterogeneity of the clusters. One could \textit{a priori} assume that the opinions of MEPs do not change much when considering documents related to the same subdomain, and therefore expect the clusters to be somewhat homogeneous regarding this aspect, or at least to see all documents related to a given subdomain gathered in the same cluster. However, the alluvial diagram shows that this is not the case at all: each cluster contains several subdomains, and certain subdomains appear in several clusters. We can think of several reasons for this. First, the subdomains do not completely encompass the characteristics of the documents, and there are other factors to take into account. We would need to detect the sentiment (in the Natural Language Processing sense, i.e. positive or negative opinion) conveyed by the document textual content to explore further this issue. Second, the dataset contains all available amendments: an amendment is generally a small modification of an existing document and can therefore lead to an easy consensus, which can introduce a bias toward this type of pattern. Third, the distribution of documents over the different subdomains is very unbalanced, and some rare subdomains are actually completely included in a single cluster, whereas other ones are widespread. For instance, themes mostly related to the economical aspects of agriculture (such as CAP mechanisms, social and structural measures, and multiple market organizations~--~see Table~\ref{tab:themes-list} for details) appear in every cluster. Of course, the clusters mechanically become purer when $k$ takes much larger values, as their sizes decrease.

Another important property of our clusterings is that, even though we do not use a hierarchical clustering method, they exhibit a quasi-hierarchical organization. More precisely, all clusters obtained for a given $k$ are kept for $k+1$, except one which is split in two to get an additional cluster. For instance, when considering $k=2$ and $k=3$: $Fr$-$k2$-$clu1$ is kept as $Fr$-$k3$-$clu1$, whereas $Fr$-$k2$-$clu2$ is split into $Fr$-$k3$-$clu2$ and $Fr$-$k3$-$clu3$. 
Interestingly, among the two clusters resulting from such a split, one always exhibits a characteristic pattern identical or very similar to that of its ancestor (e.g. $Fr$-$k3$-$clu3$ ), whereas the other's is different (e.g. $Fr$-$k3$-$clu3$). For $k>5$ though, the characteristic patterns obtained for the new clusters are not different enough to present any interest, in terms of interpretation. In the end, if the highest Silhouette value is obtained for $k=3$, the second best is $k=5$, and it additionally leads to a larger number of sufficiently different characteristic patterns. As mentioned in Section~\ref{sec:PerformingClustering}, when identifying the best $k$, it is important to consider qualitative aspects in addition to the Silhouette. Under these terms, we identify $k=5$ as our best trade-off, and discuss it in the rest of this subsection. 

\subsubsection{Characteristic Patterns}
\label{sec:ResultsFrPatterns}
In the following, we discuss each cluster obtained for $k=5$ and its corresponding pattern. 

\paragraph{Unanimity}
For space matters, Cluster $Fr$-$k5$-$clu1$ is represented separately in Figure~\ref{fig:appendix-fr-k5-clu1}, as it corresponds to a unanimity situation. Indeed, it contains a single faction, and only one negative link, between P. Le Hyaric (GUE-NGL) and M. Le Pen (NI). The emergence of such a high level of agreement was completely hidden when considering only the integrated network, and therefore could not be detected by Arinik \textit{et al}. It is the largest cluster with $100/232$ roll-calls ($43\%$), so we can assume it represents the regular voting behavior in the considered context. All the other clusters correspond to patterns containing varying antagonistic factions. This is consistent with the fact that our clusters are supposed to correspond, by construction, to distinct patterns. 

Although the unanimity case could be considered of less importance in terms of characteristic pattern, it is worth illustrating its occurrence in this context. One such roll-call is related to the improvement of applications for the protection of a designation of origin or a geographical indication (e.g. wine). Specifically, this text is about determining more explicitly the eligibility requirements to make such an application, and giving the concerned member state the responsibility to verify it. 



\paragraph{Conservatives vs. All}
The characteristic pattern associated to $Fr$-$k5$-$clu2$, shown in Figure~\ref{fig:fr-k5-clu2}, opposes the right-wing conservative group (EPP) to the rest of the MEPs, while both Euroskeptic groups (EFD and NI) abstain. This cluster contains $34/232$ ($15\%$) roll-calls. An examination of the content of the corresponding legislative documents, as well as of certain positioning documents produced by the EP groups, such as election manifestos and public letters, reveals that this voting behavior corresponds to EPP trying to block radical changes related to the CAP. These changes, as well as their blocking by the right-wing conservatives, are confirmed in a positioning paper published by S\&D about the 2013 CAP reform~\cite{SD:Manifesto}. 

Among them, one of the most important was the capping of \textit{direct payments} to farmers. Direct payments constitute a form of basic income conditioned on the implementation of certain EU rules. They represent a consequent budgetary item: $72\%$ of the EU farm budget~\cite{DirectPayments}. As such, they are difficult to reform, but the EP was willing to do so at this time, according to the chair of the AGRI Committee~\cite{DeCastroVideo}. Most roll-calls related to this topic consequently correspond to amendments to the text proposed by the commission. For instance, the first amendment proposed by S\&D, which matches the characteristic pattern, aims at capping the direct payment at $200$ k\euro. Its goal here is to decrease the support granted to large agriculture structures without affecting small- and middle-sized businesses. This change is first rejected by EPP, but a compromise is later found: it consists in raising the cap to $300$ k\euro{}.
The corresponding roll-call belongs to cluster $Fr$-$k5$-$clu3$ (discussed next), which therefore exhibits a much different characteristic pattern. 

\paragraph{Environmentalists vs. All}
Cluster $Fr$-$k5$-$clu3$, shown in Figure~\ref{fig:fr-k5-clu3}, contains $74/232$ roll-calls ($32\%$). Its characteristic pattern opposes the environmentalist group (G-EFA) to a large faction constituted of the rest of the MEPs. The far-left group (GUE-NGL) is apart, as one MEP  agrees with the environmentalists whereas the rest of his group abstains. This is very similar to the pattern obtained by Arinik \textit{et al}. when solving CC, except for the NI group and a few MEPs. In particular, Corinne Lepage, which was described by Arinik \textit{et al}. as an environmentalist member of ALDE, is placed indeed in the G-EFA faction by our method. The relevance of this faction is confirmed by her activity both at the EP, where she is very active on issues such as \textit{food safety}, and outside, through her own initiative reports and consciousness-raising conferences~\cite{ALDE2013LepageZanoniConf}. The roll-calls composing this cluster are mainly associated to amendments related to environmental aspects of agriculture, and most are proposed by G-EFA, sometimes in collaboration with C. Lepage.

Obviously, the singular position of G-EFA in this characteristic pattern is caused by its systematic opposition to the other groups on the texts associated to this cluster. However, one can distinguish two different situations. On the one hand, G-EFA tables amendments to enhance and complete the social and/or environmental regulations proposed in the amended text, and then vote in their favor. For instance, a legislative text was presented to include \textit{crop diversification} (by opposition to monoculture) among the rules that farmers must enforce to obtain direct payments. G-EFA proposed to add \textit{crop rotation} (growing different crops on the same land each year) to these requirements, as it considers it to be as crucial (in addition to crop diversification), because it helps increasing productivity as well as reducing the use of chemical fertilizer~\cite{GEFA:Manifesto12}. 

On the other hand, G-EFA opposes amendments considered as not environmentally and/or socially progressive enough, like in the case of milk quotas, for instance. An amendment was proposed to trigger milk quotas only in case of severe market imbalance. These quotas were introduced in the EU in 1984, in order to prevent milk overproduction~\cite{MilkQuota}. In 2008, the EP had already decided to increase gradually the milk quotas and abolish the quota regime in 2015. According to G-EFA, the quotas should not be abolished at all, as they favor high-quality production rather than overproducing and selling EU surplus to other countries~\cite{EP:Speeches}. Therefore, the group opposed the amendment.

By comparison, all the other groups are in favor of activating milk quotas only in case of severe market disturbance. Nonetheless, they are positioning differently on the quota system itself. For S\&D, it allows balancing between supply and demand, and contributes to market efficiency. The group assumes that its complete removal would pose serious problems. On the contrary, ALDE and EPP aim at increasing competitiveness in order to meet the global demand, and want to reduce quotas as much as possible~\cite{EP:Speeches}. Interestingly, this characteristic pattern is the only one, with unanimity, for which the Euroskeptics (NI, EFD) do \textit{not} abstain. It is difficult to assess why exactly it is the case, but this is likely to denote the importance they give to the concerned issues (yet, they have an official position on other topics, as shown in Table~\ref{tab:group-positioning}, and nevertheless abstained for some of them). Like G-EFA, they too express their will to support product quality and producer price guarantee. Moreover, it is worth noticing that this period corresponds to a significant broadening of their electoral base in rural areas~\cite{FN2014press}. Therefore, one reason for their behavior could be to support these small farmers.



\paragraph{S\&D/EPP vs. the Rest}
Cluster $Fr$-$k5$-$clu4$, shown in Figure~\ref{fig:fr-k5-clu4}, represents $18/232$ ($8\%$) roll-calls. Its characteristic pattern contains a faction formed by the far-left, environmentalist and liberal groups (GUE-NGL, G-EFA, ALDE), vs. another faction containing the socialists and conservatives (S\&D, EPP), while both Euroskeptical groups form an abstentionist faction. This constitutes a new type of pattern, different from all the others met until now, including in the baseline. In particular, it is is worth noticing that S\&D and ALDE do not belong to the same faction. Thus, if these groups alternatively side with left- and right-wing groups, as already assumed by Arinik \textit{et al}. (and as illustrated before), our method shows that they do not always do so simultaneously. 

The example of the gradual elimination of \textit{export refunds}~\cite{ExportRefunds}, and its impacts on developing countries, is particularly illustrative of the positions adopted by the different groups. Export refunds are subsidies granted for certain products (e.g. cereals, rice, sugar, beef and veal, milk and dairy products, etc.) that are exported outside the EU, and they aim at enabling EU exporters to better compete on world markets. According to ALDE, for which agriculture should be liberalized as much as possible, keeping export subsidies would cause unfair competition and might deteriorate rural development
~\cite{ALDE:Manifesto}. As a result, ALDE was in favor of phasing-out export refunds. 
G-EFA was also against export subsidies, but for different reasons. The group normally supports any market regulation guaranteeing fair producer prices and encouraging product quality. However, it considers that this specific piece of regulation goes in the opposite direction of its ideology~\cite{GEFA:PublicSpeech12}
, and therefore opposes it. S\&D also criticizes the continuation of export subsidies. But, at the same time, the group is not in favor of eliminating them, because they might weaken the EU's hand in worldwide trade negotiations~\cite{EP:Speeches}. EPP is not against eliminating the export subsidies in general, but they want to keep them in case of crisis.

\paragraph{Unholy Alliance}
For $Fr$-$k5$-$clu5$, as illustrated in Figure~\ref{fig:fr-k5-clu5}, the characteristic pattern opposes a faction gathering environmentalists and right-wing liberals and conservatives (G-EFA, ALDE, EPP), to a faction composed of the far-left and socialist groups (GUE-NGL and S\&D), while the Euroskeptics abstain once again. These factions are surprising from a political standpoint, as they exhibit a somewhat unholy alliance between environmentalists and conservatives, whose views generally clash for AGRI matters. But the pattern is also surprising when considering Arinik \textit{et al}.'s results, as they do not detect this alliance at all. The cluster contains only $6$ roll-calls ($2\%$), which shows that this situation does not happen often. 

Examining the concerned documents and debates reveals that these groups vote similarly (in this specific context), but for different reasons, as shown later by the subsequent amendments they tabled. Let us take for instance the case of \textit{green payments}~\cite{Greening}, i.e. direct payments specifically targeting agricultural practices beneficial for the climate and the environment. On the one hand, G-EFA considers that the constraints related to biodiversity are not strong enough, as already mentioned for crops when discussing $Fr$-$k5$-$clu3$. On the other hand, EPP thinks that there are too many requirements to be labeled as organic farming, and that a subset of these constraints would suffice\footnote{Incidentally, it is worth noticing that changing the definition of organic farming would make it easier to receive green payments.} Later, each group proposed a few amendments aiming at pulling the original text in its own ideological direction, regarding the points highlighted above. Each group supported its own amendments and voted against the other's, fitting the previously discussed \textit{Environmentalists vs. All} and \textit{Conservatives vs. All.} patterns, and thus highlighting their disagreement. This specific example is important, as it shows that voting similarly is not equivalent to having the same opinion. The vote similarity network is just a model of the voting process, and this example confirms that further analysis is necessary to get an accurate interpretation.

\paragraph{Comparison with the Baseline}
Our results confirm in a more objective way the assumption of Arinik \textit{et al}., based on the RCC pattern from Figure~\ref{fig:fr-all-graphs}, and according to which S\&D and ALDE sometimes vote like the left-wing groups (as in $Fr$-$k5$-$clu2$) and sometimes like the right-wing ones ($Fr$-$k5$-$clu3$). Our method additionally identifies the documents for which the EP adopts these two patterns: it turns out most of them are amendments to the same legislative propositions, in both clusters $Fr$-$k5$-$clu2$ and $Fr$-$k5$-$clu3$. 
But our method also shows that these two groups vote differently in a number of occasions ($Fr$-$k5$-$clu4$ and $Fr$-$k5$-$clu5$), a fact overlooked when using the traditional approach.

In addition, our results uncover the fact that the Euroskeptics systematically abstain on most documents, and only vote for a specific subset corresponding to the \textit{Green vs. All} pattern. This specific behavior put them apart from the rest of the groups, and maybe this is why they had been categorized by Arinik \textit{et al}. as an intermediate group, like S\&D and ALDE. However, our results show that this is an artifact of the Euroskeptics' abstentionist behavior, and that they hold a completely different position than S\&D and ALDE.

Finally, our method allows identifying the \textit{Unholy Alliance} pattern, which had completely been overlooked by Arinik \textit{et al}. It corresponds to a very surprising coalition, politically speaking, which emerged when voting for a very specific set of legislative documents, also identified by our method. By leveraging amendments related to the concerned documents, our method even allows identifying specific points of agreement and disagreement in these texts.




\subsection{Interpretation of a specific case: Italy}
\label{sec:ResultsIt}
The organization of this section is similar to the previous one: we first summarize the results of Arinik \textit{et al}. (Section~\ref{sec:ResultsItArinik}), before turning to the description of our clustering results (Section~\ref{sec:ResultsItClustering}) and the discussion of the corresponding characteristic patterns (Section~\ref{sec:ResultsItPatterns}).

\subsubsection{Baseline}
\label{sec:ResultsItArinik}
Compared to France, for the considered term, all $3$ Euroskeptical groups were represented (there was no French MEP belonging to ECR), constituting a larger proportion of the MEPs. Moreover, there were no Italian MEPs belonging to the G-EFL or GUE-NGL groups. Due to this absence of any far left and environmentalist representation, we expect to find only a small number of characteristic patterns. 

\begin{sidewaysfigure}
    \centering
    \begin{subfigure}[b]{6.8cm}
	    \centering
        \includegraphics[width=1.1\textwidth]{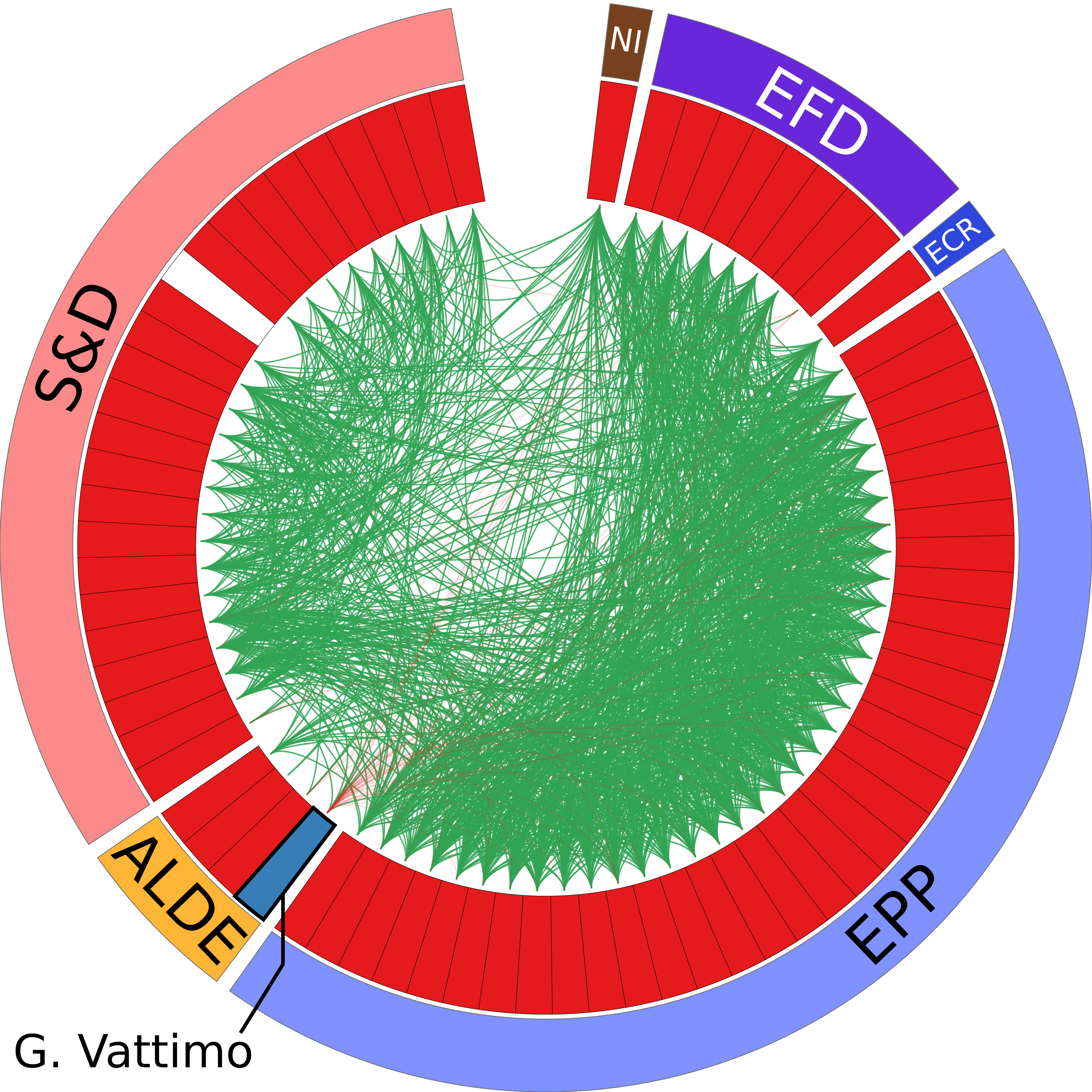}
        \caption{CC (Arinik \textit{et al}.)}
        \label{fig:CC-Italy-iknow}
    \end{subfigure}
    \hspace{1cm} 
    \vrule{}
    \hspace{0.05cm} 
    \begin{subfigure}[b]{6.8cm}
	    \centering
        \includegraphics[width=1.1\textwidth]{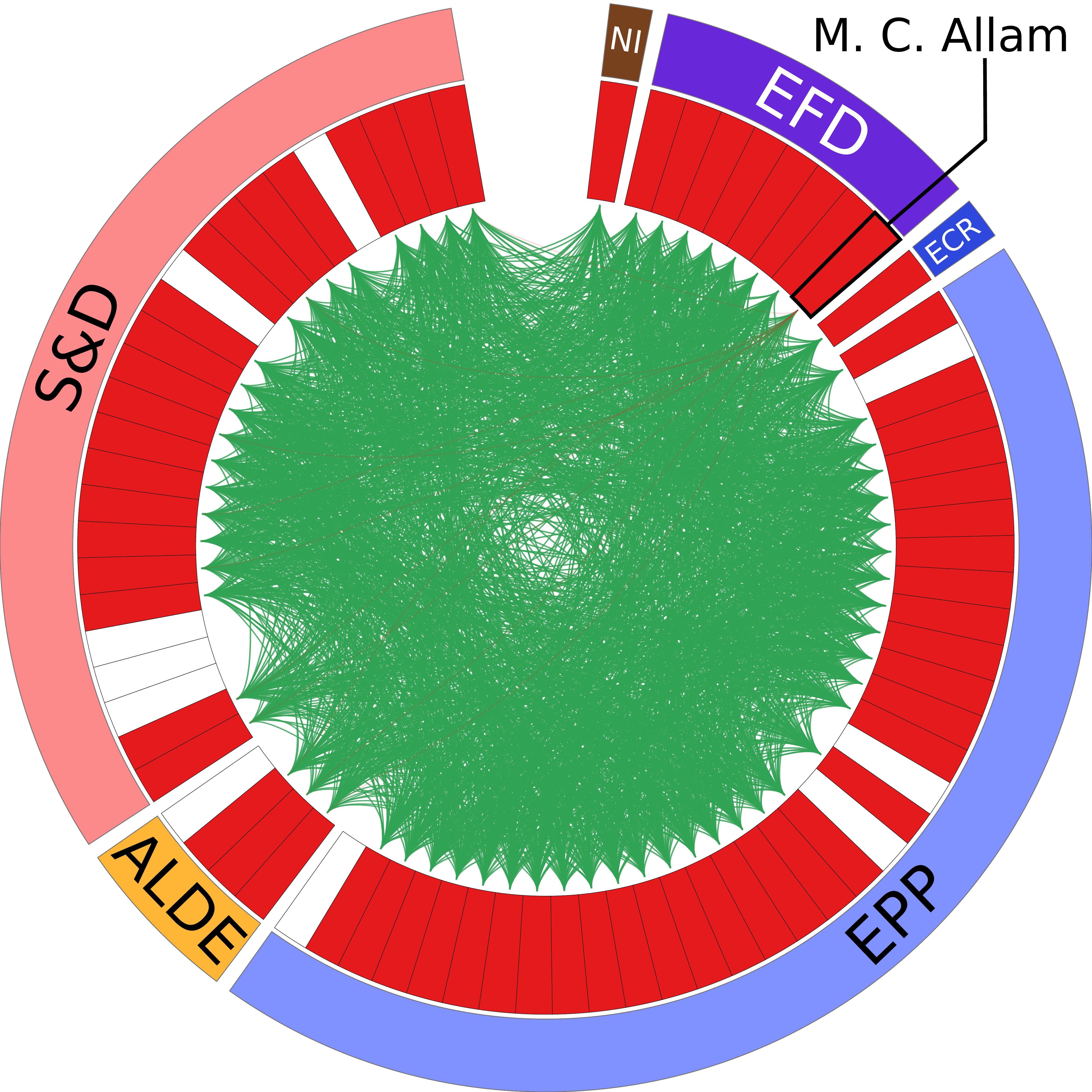}
        \caption{$It$-$k4$-$clu1$}
        \label{fig:it-k4-clu1}
    \end{subfigure}
    \hspace{1cm} 
    \begin{subfigure}[b]{6.8cm}
	    \centering
        \includegraphics[width=1.1\textwidth]{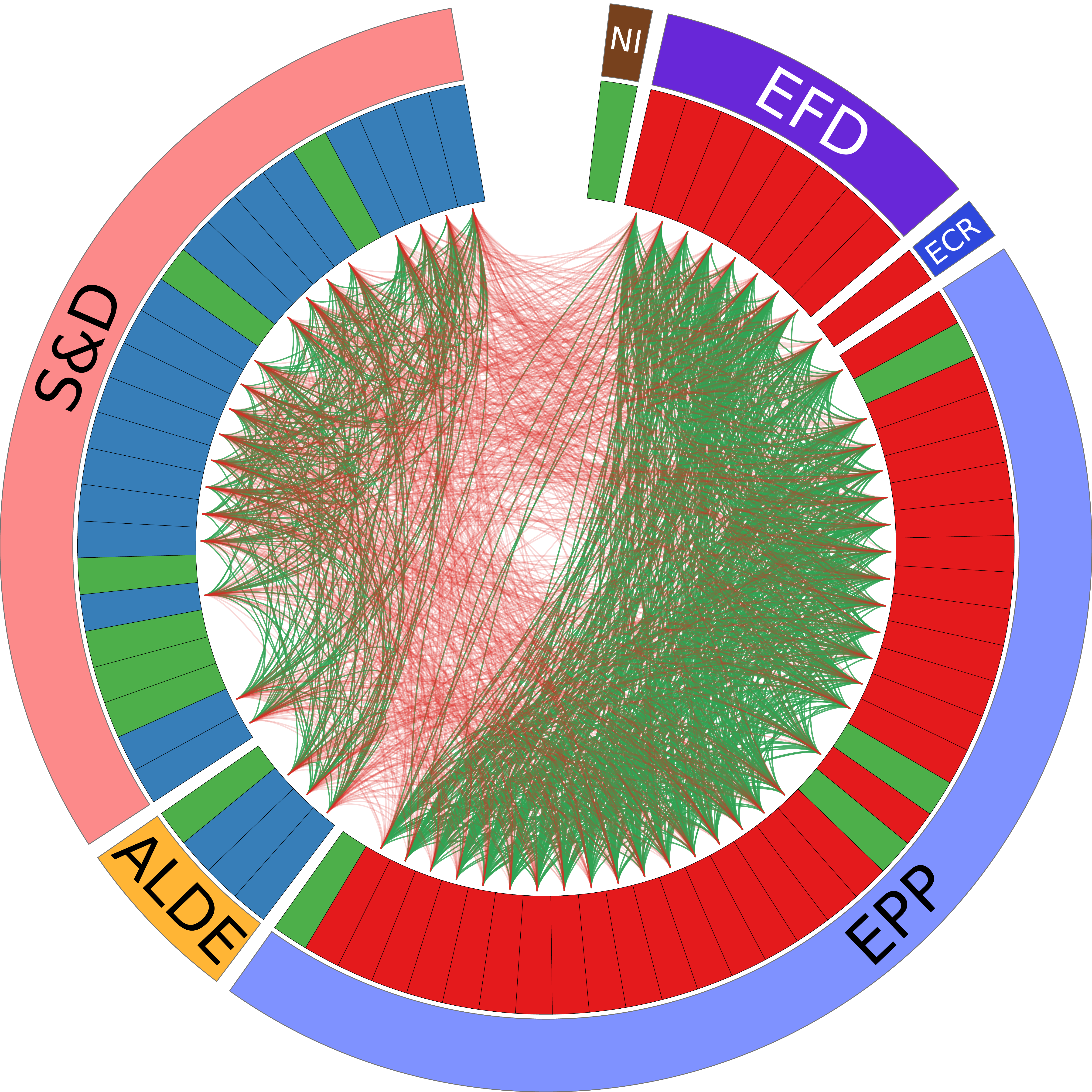}
        \caption{$It$-$k4$-$clu2$}
        \label{fig:it-k4-clu2}
    \end{subfigure}
    \\
    \begin{subfigure}[b]{6.8cm}
	    \centering
        \includegraphics[width=1.1\textwidth]{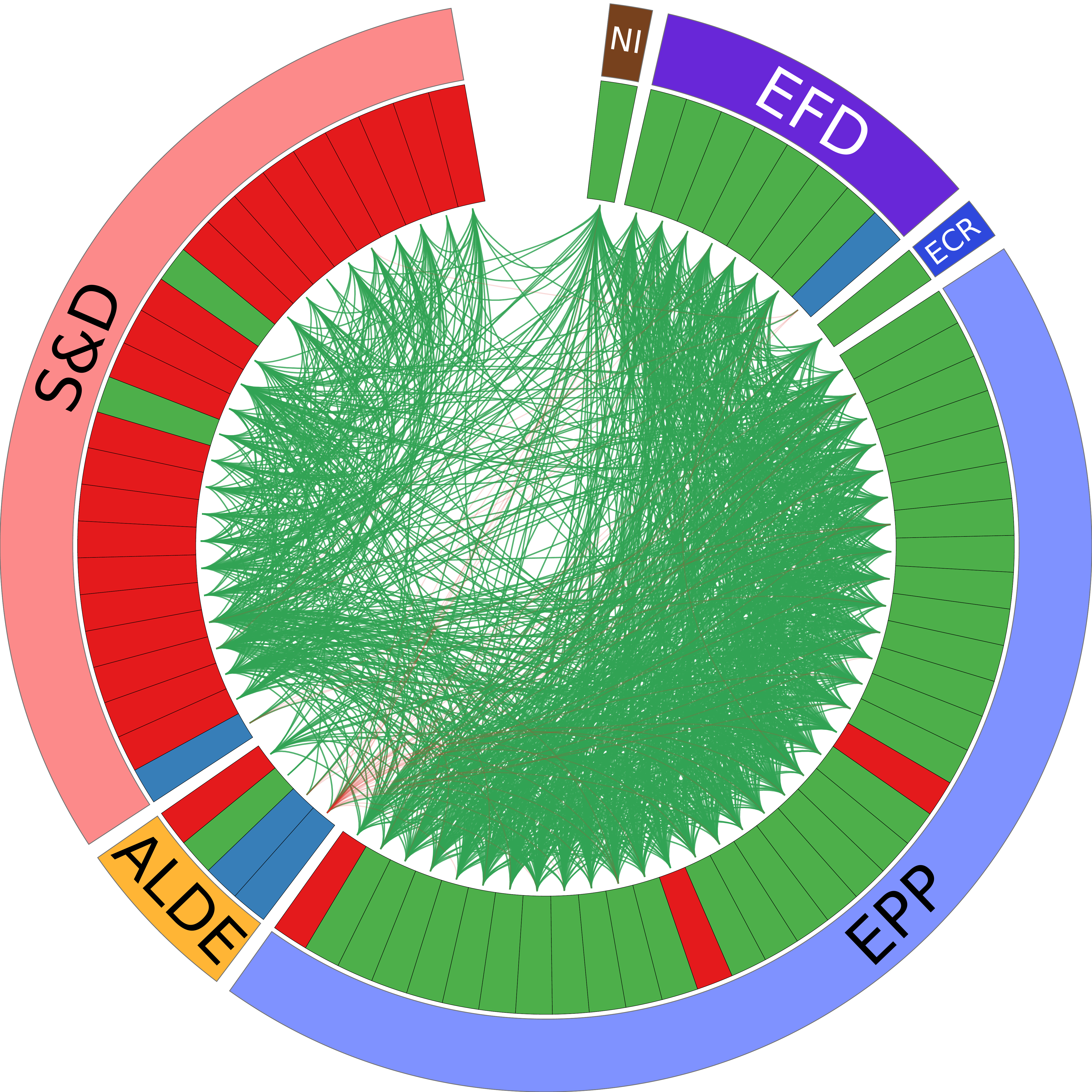}
        \caption{RCC (Arinik \textit{et al}.)}
        \label{fig:RCC-Italy-iknow}
    \end{subfigure}
    \hspace{1cm} 
    \vrule{}
    \hspace{0.05cm} 
    \begin{subfigure}[b]{6.8cm}
	    \centering
        \includegraphics[width=1.1\textwidth]{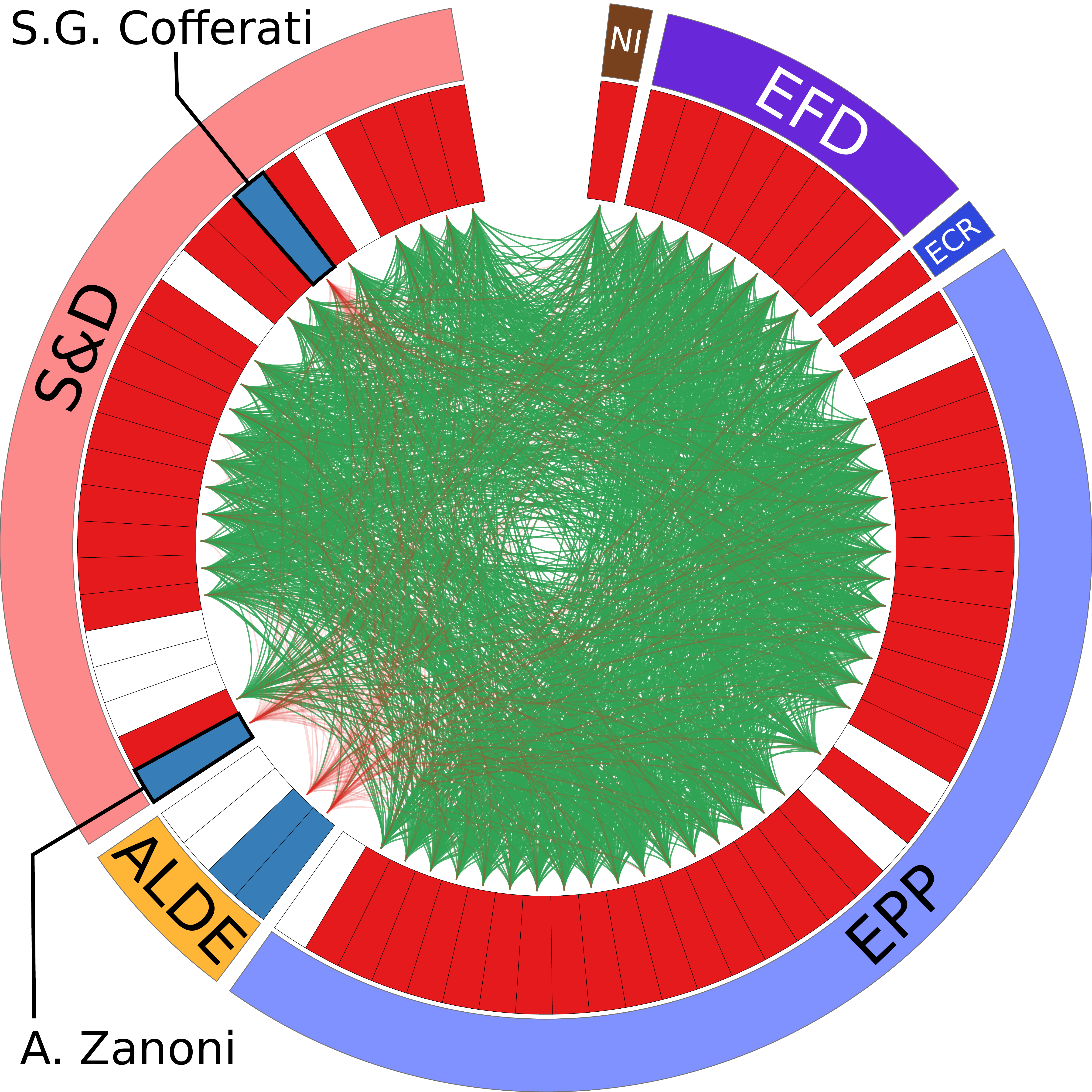}
        \caption{$It$-$k4$-$clu3$}
        \label{fig:it-k4-clu3}
    \end{subfigure}
    \hspace{1cm} 
    \begin{subfigure}[b]{6.8cm}
	    \centering
        \includegraphics[width=1.1\textwidth]{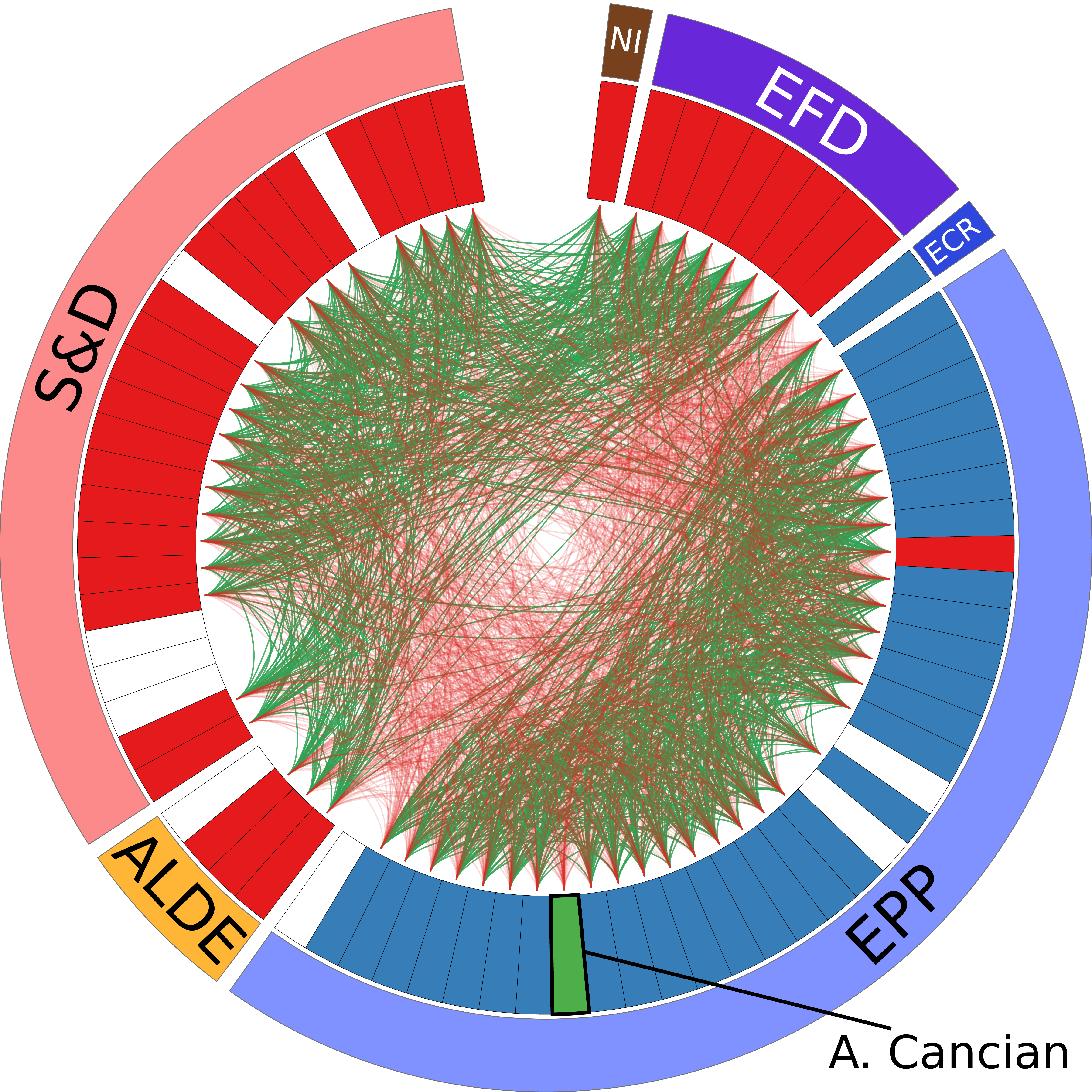}
        \caption{$It$-$k4$-$clu4$}
        \label{fig:it-k4-clu4}
    \end{subfigure}
    \caption{Voting behavior patterns of the Italian MEPs on AGRI questions in 2012-13. Red and green lines at the center represent negative and positive links, respectively (red links are drawn on top of green ones in order to improve readability). Around the links, each MEP is represented by a colored tile, whose color corresponds to the MEP's faction in the displayed pattern. The green faction in plot (f) corresponds to abstentionists. The MEPs' names are indicated separately, in Figure~~\ref{fig:appendix-fr-it-graph-legend}. The outer ring represents the political groups at the EP. The left plots show the patterns obtained by  Arinik \textit{et al}. \cite{Arinik2017} on the same integrated network when solving (a) CC and (d) RCC. The right plots show the clusters obtained with our proposed method for $k=4$: (b) $It$-$k4$-$clu1$ (\%56 of roll-calls), (c) $It$-$k4$-$clu2$ (\%11 of roll-calls), (e) $It$-$k4$-$clu3$ (\%29 of roll-calls) and  (f) $It$-$k4$-$clu4$ (\%4 of roll-calls).}
    \label{fig:it-all-graphs}
\end{sidewaysfigure}

The results obtained in the reference work \cite{Arinik2017} are shown in Figures~\ref{fig:CC-Italy-iknow} and \ref{fig:RCC-Italy-iknow} and support this idea. The Italian integrated network does not contain as many negative links as the French one. As a consequence, the method applied to solve CC (left plot) barely detects a second faction. When dealing with RCC (right plot), there are mainly a S\&D and a right-wing factions, the third one, politically located in between, being very small. Arinik \textit{et al}. assume the absence of any environmentalist among the Italian MEPs to be an important reason for the observed lack of antagonism, when compared to the French MEPs network.



\subsubsection{Clustering}
\label{sec:ResultsItClustering}
\begin{figure}[!htb]
  \begin{tabular}{lr}
    \begin{minipage}{0.45\linewidth}
    	\includegraphics[width=1.0\linewidth]{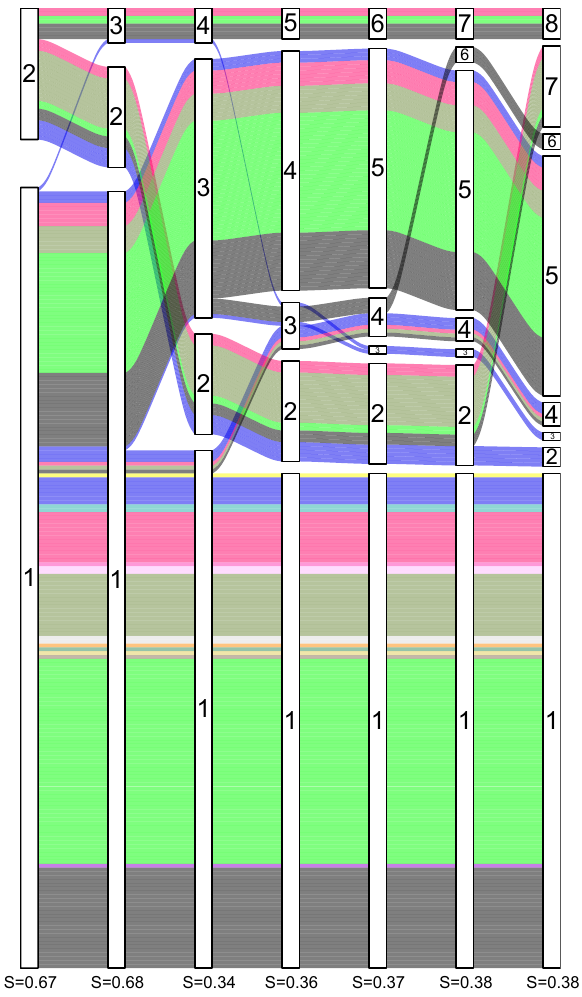}
    \end{minipage} 
    	& 
	\begin{minipage}{0.53\linewidth}
      \centering
      \begin{subfigure}[b]{1.0\textwidth}
      	 \centering
         \hspace{-0.40cm}
         \includegraphics[width=0.8\textwidth]{alluvial-legend.pdf}
      \end{subfigure}
		\begin{subfigure}[b]{1.0\textwidth}
          \hspace{-0.40cm}
          \includegraphics[width=1.0\textwidth]{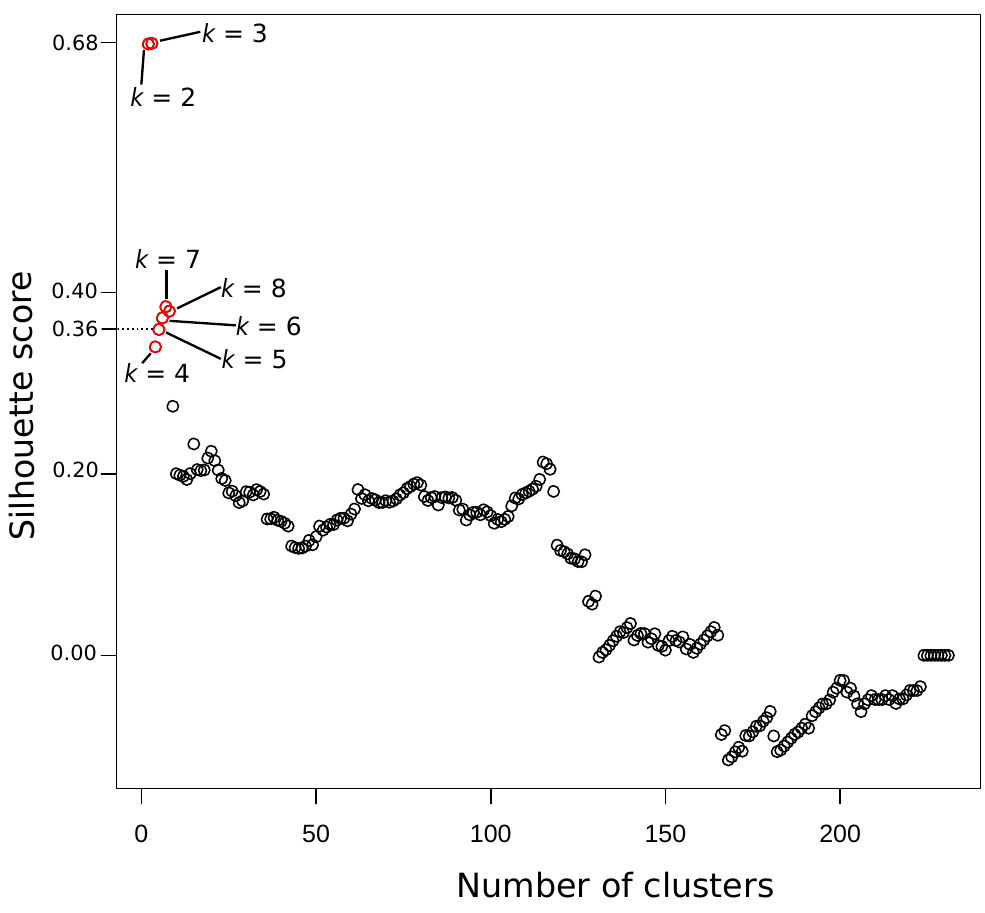}
      	\end{subfigure}
    \end{minipage}
  \end{tabular}
  \caption{Alluvial diagram (left) with its legend (top right), and Silhouette scores (bottom right) for the Italian MEPs.}
  \label{fig:it-sil-alluvial}
\end{figure}

The Silhouette plot and the alluvial diagram obtained with our method are both displayed in Figure~\ref{fig:it-sil-alluvial}. The evolution of Silhouette as a function of $k$ (bottom right plot) shows that there are only a few clusterings obtaining high Silhouette scores, even fewer than for France. They correspond to the smallest $k$ values, as for France, and as expected, but smaller ones. There is a large gap between $k=3$ and $k=4$, and an even smaller one between $k=8$ and $k=9$. 

The alluvial diagram (left plot) exhibits the same thematic heterogeneity already noted for France. It also shows that the clusters are relatively stable, and organized hierarchically, like before: for a given $k$, a newly appearing cluster is constituted by splitting a cluster obtained for $k-1$. However, an interesting difference with France is that certain new clusters are obtained by splitting the cluster associated to a quasi-unanimous voting behavior. This observation can be related to the lower level of polarity among Italian MEPs, which we previously mentioned.


Like we did for France, we do not focus on the clustering associated to the highest Silhouette, but rather on the following one, i.e. $k=4$. Indeed, its clusters have the advantage of being associated to the same characteristic patterns as those of $k=3$, while the additional cluster corresponds to unanimous votes. This cluster allows us to consider this specific situation separately. By comparison, higher $k$ values result in minor variations of the characteristic patterns, which is why we ignore them.

\subsubsection{Characteristic Patterns}
\label{sec:ResultsItPatterns}
We now discuss the clusters obtained for $k=4$ and their associated voting behavior patterns. 

\paragraph{Unanimity}
Like before, the first cluster, $It$-$k4$-$clu1$, shown in Figure~\ref{fig:it-k4-clu1}, is the largest ($130/232$ roll-calls, i.e. $56\%$), and represents the unanimity situation already observed for the French MEPs. Indeed, it contains a single faction, and a few negative links, between M. C. Allam (EFD) and a couple of S\&D and ALDE MEPs. The result obtained by Arinik \textit{et al}. when solving CC also exhibits only quasi-unanimity, which shows that, with their method, the low level of antagonism present in the activity of these MEPs was lost in the overwhelming amount of unanimity. By comparison, our approach is able to make this distinction, even if it concerns a very few roll-calls. 

The roll-calls concerned by this unanimity pattern for the Italian MEPs include all those contained in the French first cluster $Fr$-$k5$-$clu1$, which also corresponds to unanimity. However, it also contains additional roll-calls resulting in antagonistic situations among French MEPs (e.g. inclusion of permanent grassland into cross-compliance scheme~\cite{CrossCompliance}, etc.). Interestingly, the inverse is not true, as all roll-calls antagonizing Italian MEPs similarly affect French ones.

\paragraph{Quasi-unanimity}
The situation of quasi-unanimity mentioned before is identified by our method as a specific cluster\footnote{Note that this is the 3rd cluster, not the 2nd one, which is discussed later for matters of clarity.}, $It$-$k4$-$clu3$ ($67/232$ roll-calls, i.e. $29\%$), which is shown in Figure~\ref{fig:it-k4-clu3}. The associated characteristic pattern is very similar to the one obtained by Arinik \textit{et al}. when solving CC. However, it allows identifying a larger minority faction, composed of $2$ ALDE (G. Uggias, G. Vattimo) and $2$ S\&D (A. Zanoni, S.G. Cofferati) MEPs, which opposes the rest of the parliament (including some of their group fellows) on a large number of roll-calls. This result is interesting, because Arinik \textit{et al}. could detect only G. Vattimo, a philosophy professor, and explained his singular voting behavior by his atypical background and ideological position. However, our result shows he is not alone, and is therefore not really an outlier. 

A. Zanoni is an activist and important political personality in Italy, known for his actions about the protection of the environment, public health and animal welfare\footnote{\url{https://it.wikipedia.org/wiki/Andrea_Zanoni}}. In 2005, he conducted a campaign against the authorization of the largest asbestos dump in Europe, and was a local candidate for an Italian environmentalist party. A thorough inspection of his profile reveals that he was a member of ALDE for the period discussed here, and not of S\&D as indicated in our database, which explains why he votes likes this group\footnote{He switched to S\&D only later in this term. The database only contains the last affiliation recorded for the term.}. This makes A. Zanoni quite similar to C. Lepage in terms of how they handle environmental questions while belonging to the liberal group. 
Incidentally, both of them collaborated on some AGRI propositions, in particular on the topic of \textit{food safety}. S.G. Cofferati behaves similarly on these topics, although unlike A. Zanoni, he was actually a member of S\&D for the considered period. Him voting against his group is consistent with the fact that he later defended a new \textit{Left and Ecological coalition}, jointly with 9 MEPs from three different political groups (GUE-NGL, S\&D, G-EFA). All of them were members of the Progressive Caucus’ Steering Committee~\cite{LeftEcoCoalition}. 

When comparing to the French MEPs, most roll-calls contained in this cluster belong to cluster $Fr$-$k5$-$clu3$, which is associated to the \textit{Environmentalists vs. All} voting pattern. This is consistent with the environmental nature of the topics at hand (crop diversification, overproduction). Regarding the liberals, the behavior of the ALDE-dominated faction in this Italian cluster is similar to that observed for French ALDE MEPs in cluster $Fr$-$k5$-$clu4$, in which they side with G-EFA, which denotes some form of group voting discipline at the level of the EP. 

\paragraph{S\&D/ALDE vs. the Rest}
The characteristic pattern associated to the second cluster, $It$-$k4$-$clu2$ ($26/232$ roll-calls, i.e. $11\%$), illustrated in Figure~\ref{fig:it-k4-clu2}, is more classic, and opposes a faction gathering S\&D and ALDE, to a right-wing faction containing the rest of the MEPs. This highlights the fact that some roll-calls (even if a few) are very polarized, and involve very balanced groups (in terms of size), unlike what was observed for $It$-$k4$-$clu3$. This property was invisible when considering only the CC pattern of Arinik \textit{et al}. This result is more consistent with their RCC pattern, but this one does not distinguish ALDE from S\&D. Another difference is that Arinik \textit{et al}. partly doubt the presence of an actual antagonism between both factions, because most links of their network are positive. Our results are much clearer on this point. When looking at the corresponding roll-calls, we observe that the initial S\&D proposal on reduction of direct payments is also represented by this characteristic pattern. Nevertheless, contrary to the French case, the Italian S\&D and ALDE representatives apparently do not raise issues about milk quota system.


\paragraph{EPP/ECR vs. the Rest}
The characteristic pattern obtained for $It$-$k4$-$clu4$ ($9/232$ roll-calls, i.e. $4\%$), shown in Figure~\ref{fig:it-k4-clu4}, is quite similar to the previous one, but with an important difference: two Euroskeptic groups (NI and EFD) switch to the faction previously formed by the socialists and liberals (S\&D, ALDE). EPP member A. Cancian is detected as a third faction because of his frequent abstention specifically for these texts. Yet, his written declarations on the CAP 2013 reform reveal that overall, he is in line with his group. In practice, it appears that it is not always the case, as he seems to voice his hesitations on certain subjects (e.g. green payment) by voting \textsc{Abstain}.

The roll-calls forming this cluster mostly appears in the French $Fr$-$k5$-$clu4$ cluster (S\&D/EPP vs. the Rest). One example is the topic of establishing strategic stock for livestock in case of severe market disturbance. The political line of the Euroskeptic groups is to insist on such regulations aiming at preventing price volatility, which we assume to be the reason for their change of faction. Like with $It$-$k4$-$clu3$, a few roll-calls (e.g. green payment) belonging to this Italian cluster are associated to a different voting pattern for French MEPs. It is $Fr$-$k5$-$clu5$, i.e. what we called the \textit{Unholy Alliance} (G-EFA/EPP vs. the Rest). The difference in terms of voting pattern is that there are no Italian G-EFA MEPs, and that on these texts, ECR (far right-wing group) sides with EPP. 

For the interested reader, Table~\ref{tab:differences-key-votes} shows the voting behavior patterns exhibited by the French and Italian MEPs for a selection of $10$ key amendments, including those discussed in this section. To conclude, we can add that our method allows detecting subdomains treated differently by the French and Italian MEPs. For instance, the general questions related to the \textit{Agricultural Structural Fund} subdomain (represented in light blue in the alluvial diagrams from Figures~\ref{fig:fr-sil-alluvial} and \ref{fig:it-sil-alluvial}) are all placed in clusters representing unanimity situations, for both countries. On the contrary, all texts related to the \textit{Processing and marketing of agricultural products} subdomain (in pink) are treated unanimously by Italian MEPs, but not by French MEPs (presence of a G-EFA faction, see $Fr$-$k4$-$clu3$ in Figure~\ref{fig:fr-k5-clu3}).


\section{Conclusion}
\label{sec:Conclusion}
In this article, we have presented a new method to partition multiplex signed networks. It is based on constraints arising from a real-world application: the analysis of voting data at parliaments and similar institutions. We want not only to detect antagonistic factions of voters, as in most works of the literature, but also to identify characteristic voting patterns and the circumstances (bills) leading to their appearance. For this purpose, we define a multiplex signed graph, whose layers correspond to roll-calls. Our four-stepped method first partitions each layer separately, computes the similarity between these partitions, then groups them using cluster analysis, and characterizes each such group. By applying it to a subset of the 7\textsuperscript{th} term European Parliament dataset presented in \cite{Arinik2017}, we could confirm some of the assumptions made by Arinik \textit{et al}. based on a traditional approach. For instance, we could show that the French S\&D and ALDE MEPs alternatively side with the left- and right-wing MEPs on certain questions. But we could also uncover some overlooked points, such as the fact that despite the absence of any environmentalist among the Italian MEPs, a certain number of documents cause a strong antagonism.

On the one hand, by comparison to existing approaches, our method has the following advantages. First, it undergoes much less of the information loss appearing when integrating the raw voting data to extract the voting similarity networks. Second, in addition to antagonistic factions of voters, it allows identifying groups of legislative propositions causing the same voting pattern for these factions. Third, it does not require to filter out (quasi-)unanimous propositions, or to discard week links appearing in the model. Fourth, it explicitly represents abstention, which allows detecting relevant groups of abstentionists. On the other hand, the main limitation is that, because this was not a relevant issue for the data considered in this article, our approach does not model time explicitly. 
Another issue with our work is that, as our main objective was to use these EP data to illustrate the benefits of our method, a proper interpretation of our results would require further work.

Our work could be extended in several ways. First, by applying our method more systematically to the whole EP dataset. This could cause some computational issues when solving the Correlation Clustering problem, but these could be solved by using an approximate method in lieu of Ex-CC \cite{Levorato2017a}. Second, we observed that when applying $k$-medoids with an increasing $k$ value, the obtained series of clusterings is almost hierarchical: it might be more appropriate to directly apply a hierarchical clustering algorithm instead. From the application perspective, we have noticed that the subdomains used to characterize the documents are not enough to interpret the detected clusters. A sentiment analysis of the documents textual content could provide the missing information, but this opens a number of technical difficulties. Finally, our method is generic, in the sense it could be applied to any system with similar properties, so this could constitute another perspective. For example in the context of document/artwork classification, the opinion expressed by a selection of specialists can produce a set of behavior patterns, each one representing an individual opinion \cite{Bansal2002}. Our method could be used to group specialists that share the same point of view. As another example, signed networks can also be used to represent lipread consonants similarity \cite{Manning1991}. Some studies involve a set of individuals rating (on a given scale) the subjective similarity between pairs of consonants, and this can be considered as the patterns in our framework. Our method could provide a classification of the set of individuals according to their sensibilities, origins or disabilities. 



\paragraph{Acknowledgments.} This research benefited from the support of the Agorantic FR3621, as well as the FMJH Program PGMO and from the support to this program from EDF-THALES-ORANGE-CRITEO.

\phantomsection\addcontentsline{toc}{section}{References}
\printbibliography






\appendix

\section{EP- and CAP-Related Concepts}
\label{sec:appendix-definitions}
This appendix aims at providing the reader with some context regarding the European Parliament and the Common Agricultural Policy, through the definition of the main related concepts.


\paragraph{Common Agricultural Policy} The CAP has been established in 1957 to help EU farmers face agricultural challenges, which can be social (low labor force, etc.), economical (increase of inequalities, subsidies, etc.) or environmental (water pollution, global warming, etc.)~\cite{EPCouncilCAP}. 

The CAP is funded and shared by all the members state of the European Union. It currently takes action with income support (i.e. income stability through direct payments), market measures (i.e. dealing  with difficult market situations such as severe market imbalance, etc.) and rural development measures (specific challenges facing rural areas).

\paragraph{Pillars I and II} The CAP has two main principles called pillars in the EU jargon~\cite{AgriGlossary}. The \textit{Pillar I} is support to farmers' incomes. It is established through direct payments and market measures, and is entirely financed from the European Agricultural Guarantee Fund (EAGF). \textit{Pillar II} is support to the development of rural areas. It is established through Rural Development programmes, and is co-financed by the European Agricultural Fund for Rural Development (EAFRD).

\paragraph{Direct payments} Direct payments were introduced in 1992 to support the incomes of farmers~\cite{AgriGlossary}. Before that, the CAP was supporting the prices instead (an indirect form of payment). 

\paragraph{Green payment} A specific form of direct payment, granted for agricultural practices beneficial for the climate and the environment. Member states must allocate 30\% of their direct payment allocation to this greening payment~\cite{Greening}. After the 2013 CAP reform, the three greening obligations are: 1) Crop diversification, 2) Maintenance of permanent grassland, 3) Ecological focus area.


\paragraph{Co-financing scheme} Member states have a certain discretion regarding the final design of Rural Development measures. The support granted under each measure is shared between the EU and the concerned member state. This arrangement is known as co-financing~\cite{AgriGlossary}.

\paragraph{Milk quota system} A quota system designed to control the production of dairy products in the EU. Its aim was to avoid overproduction. It was first established in 1983, after the occurrence of large volume of milk (and other dairy products) surplus, which had to be bought by European Commission~\cite{MilkQuota}.

\paragraph{Export subsidy} Export subsidies (or \textit{export refunds}) are special payments provided by the government to encourage export of goods, and to discourage sale of goods on the domestic market~\cite{AgriGlossary}. 

\paragraph{Cross-compliance scheme} The cross-compliance scheme is a set of requirements (e.g. food safety, animal health, plant health, etc.) that farmers should satisfy in order to receive direct payments~\cite{AgriGlossary}.


\paragraph{Permanent grassland} It corresponds to a land used permanently (usually more than five consecutive years)~\cite{AgriGlossary}. 

\paragraph{Natura 2000} Natura 2000 is a network of nature protection areas in the European Union. The aim is to protect the most seriously threatened habitats and species across Europe~\cite{AgriGlossary}.

\paragraph{Water Framework Directive (WFD)} This legislation regarding water protection aim for the future~\cite{AgriGlossary}.



\section{Key Elements of the 2013 CAP Reforms}
\label{sec:appendix-reforms}
The 2013 CAP reform, which covers the period from 2014 to 2020, has a particularity compared to the previous reforms, because it is the first time that the EP has a say in their adoption, by co-legislating with the Council of the EU. The reform process was launched by the EU commission April 2010 through public debates and conferences. The final adoption of the legal texts took place on December 2013. Many of the regulations took effect starting from 2015, so that member states would have enough time to prepare.


The 2013 CAP reforms can be grouped in the following $4$ different sets of regulations~\cite{EPCouncilCAP, EPCommOverviewCAP, Matthews:CAP}:
\begin{itemize}
\item Rules for direct payments to farmers:
	\begin{itemize}
      \item the \textit{greening} of farm payments, through the introduction of environmental practices, such as crop diversification, and maintaining ecologically rich landscape features and a minimum area of permanent grassland;
      \item more equality in the distribution of funds received by farmers across the EU, and a reduction in payments above a certain amount for the biggest farms;
      \item better targeting of income support to farmers (active farmers, young farmers, small farmers);
    \end{itemize}
\item A common organization of the markets in agricultural products:
	\begin{itemize}
    \item Strengthening producer power in the food chain: Contracts (contract duration, quantity and quality requirements, etc., for agricultural products), price negotiations in dairy sector (maintaining effective competition), withdrawals (implementing private supply management to fix prices under difficult situations);
    \end{itemize}
\item Support for rural development:
	\begin{itemize}
      \item Encouraging knowledge transfer and innovation in forestry and rural areas;
      \item Enhancing farm viability and competitiveness;
      \item Promoting food chain organization, including processing and marketing of agricultural products, risk management in agriculture;
      \item Promoting resource efficiency and supporting environment-friendly practices (i.e. low carbon use);
      \item Promoting social inclusion, poverty reduction.
    \end{itemize}
\item Financing, management and monitoring of the common agricultural policy:
	\begin{itemize}
      \item European Price Monitoring Tool: It aims at collecting necessary information
  about the stages of food supply chain in the EU and the Member States (e.g. exports and imports, farm gate prices, consumer prices);
      \item Food security and assessment of impact on developing countries.
	\end{itemize}
\end{itemize}

\section{Hierarchy of AGRI-related topics}
\label{sec:appendix-DataCat}
In order to characterize subgroups of legislative propositions in a topical way, we use EUR-Lex\footnote{\url{http://eur-lex.europa.eu}}, the website of the EU for the publication of official documents such as treaties and legislation. For indexing matters, this website provides a hierarchical nomenclature of topics. 

\begin{table}[htb!]
\footnotesize
	\begin{tabular}{l}
	\hline
	\textbf{Domains and subdomains for the legislative propositions related to agriculture in 2012-13}\\
	\hline
	\DTsetlength{0.2em}{0.7em}{0.2em}{0.4pt}{0pt}
	\begin{minipage}{15cm}\dirtree{%
		.1 Agriculture (AGRI).
			.2 Basic provisions (BP).
				.3 (22.9\%) \textbf{Common agricultural policy mechanisms (CAPM)}.
			.2 Agricultural structures (AS).
				.3 (15.9\%) \textbf{Social and structural measures (SSM)}.
				.3 (0.8\%) \textbf{Processing and marketing of agricultural products (PMAP)}.
			.2 Agricultural structural funds (ASF).
				.3 (0.8\%) \textbf{General (GEN)}.
				.3 (0.4\%) \textbf{European Agricultural Guarantee Fund (EAGF)}.
				.3 (10.8\%) \textbf{European Agricultural Fund for Rural Development (EAFRD)}.
			.2 Approximation of laws and health measures (ALHM).
				.3 (8.6\%) \textbf{Animal health and zootechnics (AHZ)}.
				.3 (0.4\%) \textbf{Seeds and seedlings (SS)}.
			.2 Products subject to market organization (PSMO).
				.3 (0.4\%) \textbf{Seeds (SEED)}.
				.3 (0.4\%) \textbf{Wine (WINE)}.
				.3 (37.9\%) \textbf{Arrangements covering multiple market organizations (ACMOMO)}.
		.1 Regional policy and coordination of structural instruments (REGP).
			.2 (0.8\%) \textbf{Coordination of structural instruments (CSI)}.
		.1 Environment, consumers and health protection (ENVI).
			.2 Consumers(CONS).
				.3 (0.4\%) \textbf{Protection of economic interests (PEI)}.
		.1 External relations (EXTR).
			.2 (0.4\%) \textbf{Bilateral agreements with non-member countries (BANC)}.
		.1 Fisheries (FISH).
			.2 (0.4\%) \textbf{External (EXT)}.
	}\end{minipage}
	\\
	\hline
	\end{tabular}
	\caption{EUR-Lex subdomains used to categorize the legislative propositions themes}
    \label{tab:themes-list}
\end{table}

The agriculture domain (AGRI) is represented over $4$ hierarchical levels. The fourth one is too specific: for the considered period, each document basically concerns a different subdomain, which would prevent us from detecting any relevant pattern. For this reason, we work with the third level, which is represented in Table~\ref{tab:themes-list}. The subdomains relevant for the considered time period are represented in bold. Based on the titles and summaries of the legislative propositions, we manually annotate each of them with the most appropriate subdomains. Note that one document can be associated with several subdomains. Moreover, in addition to their agriculture-related subdomain(s), some documents also belong to other \textit{domains}, such as environment (ENVI) or fisheries (FISH). We proceed similarly to identify their subdomains, and these are also shown in the table. Finally, Table~\ref{tab:themes-list} also displays the proportions of propositions described by each subdomain, for the considered selection.

\section{Positioning of EP Groups on CAP Topics}
\label{sec:appendix-Positions}
In order to interpret the groups of similarly voting MEPs identified by our method, we need some real-world reference regarding the positioning of the EP groups on agricultural questions. To this aim, we have manually reviewed the official material published by the EP groups on these topics for the 2012-13 period: manifestos, positioning papers, and revised written transcriptions of speeches made at the EP. Table~\ref{tab:group-positioning} summarizes the information that we collected. Each row corresponds to one of the main agricultural topics, and each column to an EP group. Empty cells reflect either the fact that the concerned group did not take any position relatively to the considered topic, or that no source could be found to describe this position. It appears that agreement between groups covers all the spectrum, ranging from complete disagreement (e.g. GUE-NGL and G-EFA) to almost perfect agreement (e.g. GUE-NGL and G-EFA).

\begin{sidewaystable}
	\scriptsize
	\caption{Position of the EP Groups regarding AGRI-related topics for the 2012-13 period.}
    \label{tab:group-positioning}
    \hspace{-0.5cm}
	\begin{tabular}{| C{2.5cm} | C{2cm} | C{2.5cm} | C{2.2cm} | C{2.2cm} | C{2.2cm} | C{2cm} | C{2cm} | C{2cm} |}
    	\hline
        Subject 	& GUE-NGL & G-EFA & S\&D & ALDE & EPP & EFD & ECR & NI\\
        \hline
 		1) Reduction of direct payments
        	& \textsc{For} reduction (ceiling at 100 k\euro)~\cite{GUENGL:Manifesto11}
            & \textsc{For} reduction (starting from 50 k\euro)~\cite{GEFA:PublicSpeech12}
            & \textsc{For} reduction (starting from 150 k\euro)~\cite{EP:Speeches}
            & -
            & \textsc{Against} reduction~\cite{EP:Speeches}
            & \textsc{Against} reduction~\cite{EP:Speeches}
            & \textsc{Against} reduction~\cite{EP:Speeches}
            & -
            \\
        \hline
 		2) Maintaining milk quotas
        	& \textsc{For} quotas (with flexibility)~\cite{EP:Speeches}
            & \textsc{For} quotas (food security purposes)~\cite{EP:Speeches}
            & \textsc{For} quotas~\cite{EP:Speeches}
            & \textsc{Against} quotas (competitiveness purposes); \textsc{For} transition period~\cite{ALDE:Manifesto}
            & \textsc{Against} quotas (competitiveness purposes)~\cite{EP:Speeches}
            & -
            &\textsc{Against} quotas (low food price purposes)~\cite{EP:Speeches}
            & \textsc{For} quotas (fair price purposes)~\cite{NI:LePenSpeech,NI:GollnischSpeech}
            \\\hline
 		3) Export subsidies
        	& -
            & \textsc{Against} subsidies~\cite{GEFA:PublicSpeech12,GEFA:Manifesto12}
            & \textsc{Against} subsidies~\cite{EP:Speeches}
            & \textsc{Against} subsidies (with transition period)~\cite{ALDE:Manifesto}
            & \textsc{Against} subsidies; \textsc{For} exceptional subsidies~\cite{EP:Speeches}
            & -
            & -
            & -
            \\\hline
 		4) Competitiveness
        	& \textsc{Against} current system (too competitive)~\cite{EP:Speeches}
            & \textsc{Against} current system (too competitive)~\cite{EP:Speeches}
            & -
            & \textsc{For} competitiveness~\cite{ALDE:Manifesto}
            & \textsc{For} competitiveness (better functioning of supply chain purposes)~\cite{EP:Speeches,EPP:Press13}
            & -
            & \textsc{For} competitiveness (low food price purposes)~\cite{EP:Speeches}
            & \textsc{Against} current system (fair price purposes)~\cite{NI:LePenSpeech,NI:GollnischSpeech}
            \\\hline
 		5) Aid for rural development
        	& -
            & \textsc{Against} current scheme (not enough); \textsc{For} co-financing~\cite{EP:Speeches}
            & \textsc{For} transfer from Pillar I to Pillar II; \textsc{Against} co-financing~\cite{SD:Manifesto}
            & \textsc{For} transfer from Pillar I to Pillar II, co-financing~\cite{EP:Speeches}
            & \textsc{For} transfer from Pillar I to Pillar II,  producers as market actors~\cite{EP:Speeches}
            & -
            & \textsc{Against} risk management measures (not enough)~\cite{EP:Speeches}
            & \textsc{Against} the current scheme (not enough)~\cite{NI:LePenSpeech}
            \\\hline
 		6) Food quality vs. quantity
        	& For \textsc{quality}~\cite{GUENGL:Manifesto11}
            & For \textsc{quality}~\cite{GEFA:Manifesto12,EP:Speeches}
            & -
            & For \textsc{quantity}~\cite{ALDE:Manifesto}
            & For \textsc{quantity} and \textsc{quality}~\cite{EPP:Press13}
            & For \textsc{quality}~\cite{EFD:DeVilliersSpeech}
            & For \textsc{quantity} and \textsc{quality}~\cite{EP:Speeches}
            & For \textsc{quality}~\cite{NI:GollnischSpeech}
            \\\hline
        7) Enhancing environmental measures
        	& \textsc{For} organic farming; \textsc{Against} GMO~\cite{GUENGL:Manifesto11,EP:Speeches}
            & \textsc{For} crop rotation and diversification, promoting biodiversity, organic farming, permanent grasslands, \textsc{Against} GMO, intensive agriculture\cite{GEFA:Manifesto12,GEFA:PublicSpeech12}
            & \textsc{For} reducing use of chemicals, promoting biodiversity, energy savings; \textsc{Against} intensive farming~\cite{SD:Manifesto}
            & \textsc{For} greener CAP, energy savings, tackling climate change through innovative solutions~\cite{ALDE:Manifesto}
            & -
            & \textsc{Against} GMO~\cite{EFD:DeVilliersSpeech}
            & Not the priority~\cite{EP:Speeches}
            & -
            \\
        \hline
	\end{tabular}
\end{sidewaystable}

\section{Comparison Between Italy and France}
\label{sec:appendix-Comparison}
Table~\ref{tab:differences-key-votes} shows the voting behavior patterns exhibited by the French and Italian MEPs for a selection of $10$ key amendments, including those discussed in Section~\ref{sec:Results}. An amendment treated unanimously by Italian MEPs might leads to an antagonistic situation for French MEPs, but the opposite is not true (not only for those $1$0 amendments, but also for all the amendments considered here, in general).

Our method allows detecting subdomains treated differently by the French and Italian MEPs. For instance, the general questions related to the \textit{Agricultural Structural Fund} subdomain (represented in light blue in the alluvial diagrams from Figures~\ref{fig:fr-sil-alluvial} and \ref{fig:it-sil-alluvial}) are all placed in clusters representing unanimity situations, for both countries. On the contrary, all texts related to the \textit{Processing and marketing of agricultural products} subdomain (in pink) are treated unanimously by Italian MEPs, but not by French MEPs (presence of a G-EFA faction, see the figure related to $Fr$-$k4$-$clu3$ in the article). This fact is also seen in Table~\ref{tab:differences-key-votes}. For instance, although Amendments 144 of B7-0082 and 147 of B7-0079 related to environmental aspects are treated in the same way for France ($Fr$-$k4$-$clu3$), they are treated differently for Italy.

\begin{table}[htb]
	\footnotesize
	\caption{Voting behavior patterns exhibited by the French and Italian MEPs for a selection of $10$ key amendments}
    \label{tab:differences-key-votes}
	\centering
	\begin{tabular}{| C{3cm} | C{5cm} | C{3cm} | C{3cm} |}
    	\hline
        Amendment no & Subject 	& French MEPs' positioning & Italian MEPs' positioning \\
        \hline
        Am 146 of B7-0081 & Natura 2000 and Water Framework Directive payments: foreseeing incompatible requirements in some regions & Conservatives vs. All ($Fr$-$k5$-$clu2$) & S\&D/ALDE vs. the Rest ($It$-$k4$-$clu2$) \\
        \hline
        Am 105 PC of B7-0079 & Reduction to direct payments to farmers (initial S\&D proposal) 	& Conservatives vs. All ($Fr$-$k5$-$clu2$) & S\&D/ALDE vs. the Rest ($It$-$k4$-$clu2$) \\
        \hline
        Am 109 of B7-0080 & Measures against market disturbance: strategic stocks for livestock & Conservatives vs. All ($Fr$-$k5$-$clu2$) & EPP/ECR vs. the Rest ($It$-$k4$-$clu4$) \\
 		\hline
        Am 144 of B7-0082 & Inclusion of permanent grassland into cross-compliance scheme & Environmentalists vs. All ($Fr$-$k5$-$clu3$) & Unanimity ($It$-$k4$-$clu1$) \\
         \hline
        Am 147 of B7-0079 & Crop diversification & Environmentalists vs. All ($Fr$-$k5$-$clu3$) & Quasi-unanimity ($It$-$k4$-$clu3$) \\
         \hline
         Am 67 of B7-0080 & Measures against market disturbance: milk and sugar quota 	&  Environmentalists vs. All ($Fr$-$k5$-$clu3$) & Quasi-unanimity ($It$-$k4$-$clu3$) \\
        \hline
         Am 450 of B7-0080 & Elimination of export refunds & Environmentalists vs. All ($Fr$-$k5$-$clu3$) & Quasi-unanimity ($It$-$k4$-$clu3$) \\
        \hline
         Am 471 of B7-0080 & Inclusion of export refunds for a limited period (by 2013) &  S\&D/EPP vs. the Rest ($Fr$-$k5$-$clu4$) & Unanimity ($It$-$k4$-$clu1$) \\
        \hline
         Am 223 of B7-0082 & Food security and assessment of impact on developing countries & S\&D/EPP vs. the Rest ($Fr$-$k5$-$clu4$) & Quasi-unanimity ($It$-$k4$-$clu3$) \\
          \hline
        Am 106/1 of B7-0079 & Payment of agricultural practices beneficial for the climate and the environment 	& EPP-Green/rest ($Fr$-$k5$-$clu5$) & EPP/ECR vs. the Rest ($It$-$k4$-$clu4$)\\
	\hline
    \end{tabular}
\end{table}

\section{Additional Plots}
\label{sec:appendix-fr-k5-clu1}
\begin{figure*}[htb!]
	\centering
    \includegraphics[height=6cm]{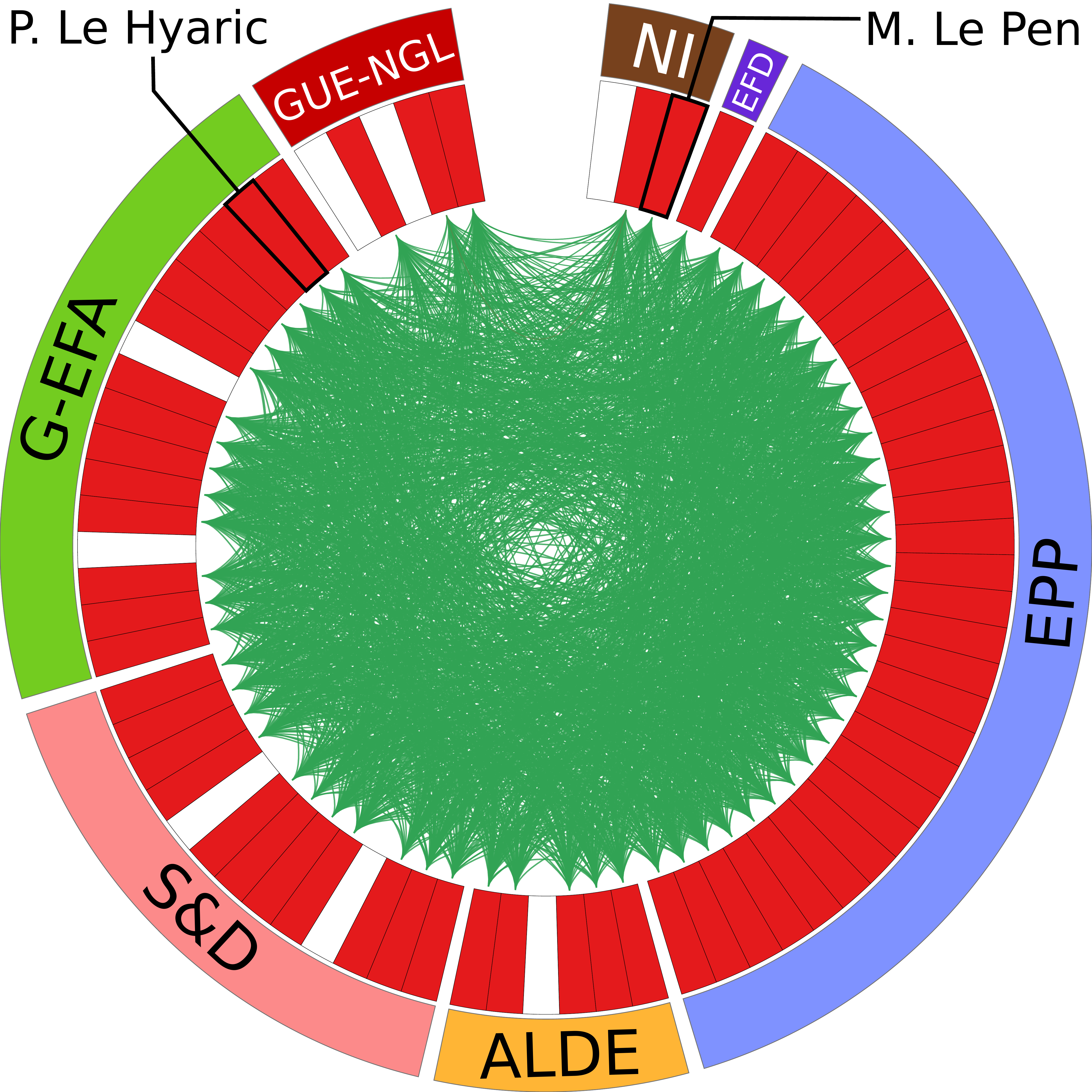}\hspace{2mm}
	\caption{Characteristic pattern of cluster $Fr$-$k5$-$clu1$, in complement to Figure~\ref{fig:fr-all-graphs}}
	\label{fig:appendix-fr-k5-clu1}
\end{figure*}

\begin{sidewaysfigure}[htb!]
	\centering
    \includegraphics[width=0.490\linewidth]{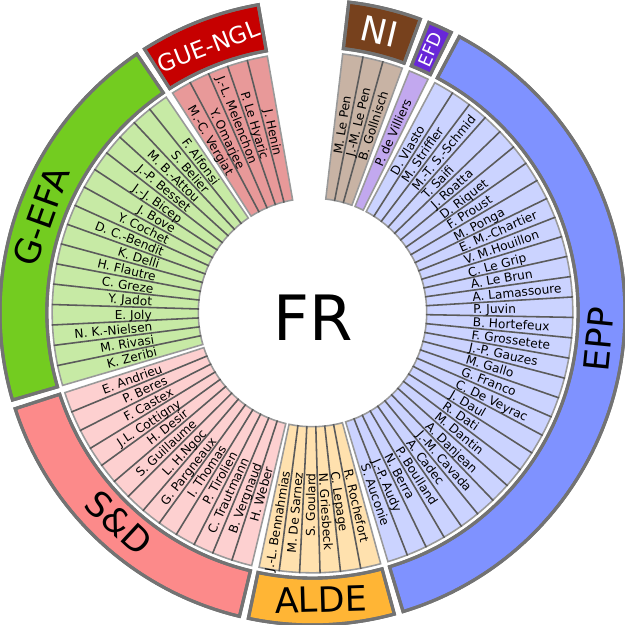}\hspace{2mm}
    \includegraphics[width=0.490\linewidth]{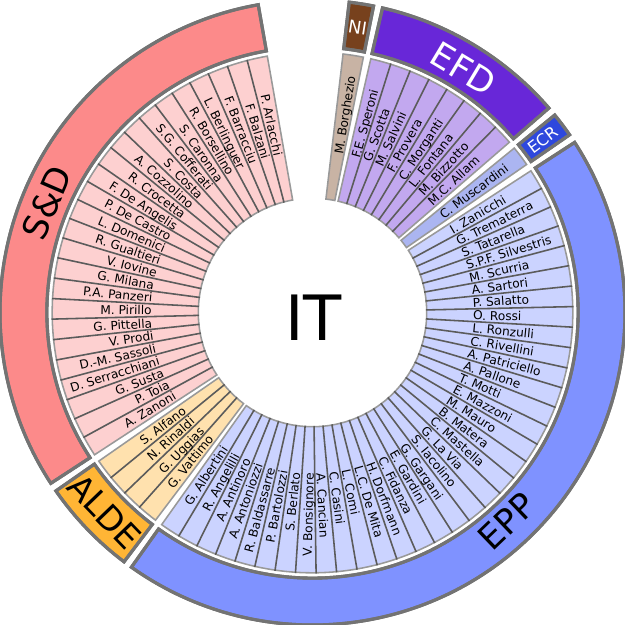}
	\caption{Name of the French (left) and Italian (right) MEPs studied in Section~\ref{sec:Results}. The layout is the same as in Figures~\ref{fig:fr-all-graphs} and \ref{fig:it-all-graphs}.}
	\label{fig:appendix-fr-it-graph-legend}
\end{sidewaysfigure}

Figure~\ref{fig:appendix-fr-k5-clu1} represents cluster $Fr$-$k5$-$clu1$, which was not included in Figure~\ref{fig:fr-all-graphs} due to lack of space. Figure~\ref{fig:appendix-fr-it-graph-legend} contains the names of all French and Italian MEPs. Their position in the figure matches the one they have in Figures~\ref{fig:fr-all-graphs} and \ref{fig:it-all-graphs}, respectively.

\end{document}